\documentclass[
    ,conference
]{IEEEtran}
\IEEEoverridecommandlockouts

\usepackage{amsmath, amsthm}
    
\usepackage{graphicx}
\usepackage[table]{xcolor} 
\usepackage{hyperref, url}
\usepackage{adjustbox}
\usepackage{booktabs, multirow}
\usepackage{caption}
\usepackage{subcaption}
\usepackage{enumitem}
\usepackage{lipsum}
\usepackage{setspace}
\usepackage[normalem]{ulem}

\usepackage{algorithm, algpseudocode}
    \definecolor{BLUE}{rgb}{.0, .2, .6}
    \definecolor{BLUEalt}{HTML}{1e50a2}
    \definecolor{RED}{HTML}{c9171e}
    \algrenewcommand{\alglinenumber}[1]{{\scriptsize\bfseries\ttfamily\color{RED}#1}}

\usepackage{thmtools}
\declaretheoremstyle[%
  spaceabove=4pt,%
  spacebelow=4pt,%
  headfont=\normalfont\itshape,%
  postheadspace=1em,%
  qed=\qedsymbol%
]{mystyle}

\usepackage{footnote}
\makesavenoteenv{tabular}
\makesavenoteenv{table}

\usepackage{xspace}

\usepackage{tikz}
\usepackage{array}
\usepackage{booktabs}
\usepackage{tabu}

\usepackage[zerostyle=d]{newtxtt}

\usepackage{cite}
\usepackage{amsmath,amssymb,amsfonts}
\usepackage{graphicx}
\usepackage{textcomp}
\usepackage{comment}
\usepackage{xcolor}
\usepackage{threeparttable}
\def\BibTeX{{\rm B\kern-.05em{\sc i\kern-.025em b}\kern-.08em
    T\kern-.1667em\lower.7ex\hbox{E}\kern-.125emX}}
\newcommand{\revise}[1]{\textcolor{black}{#1}}
    
\begin{document}

\title{Accelerating Parallel Write via Deeply Integrating Predictive Lossy Compression with HDF5}

\newcommand{\BetterMark}{$^\star$}

\author{
Sian Jin\BetterMark,
Dingwen Tao\BetterMark\thanks{Corresponding author: Dingwen Tao (\url{dingwen.tao@wsu.edu}).
},
Houjun Tang\BetterMark,
Sheng Di\IEEEauthorrefmark{2},
Suren Byna\IEEEauthorrefmark{3},
Zarija Lukic\IEEEauthorrefmark{3},
Franck Cappello\IEEEauthorrefmark{2}\\
\BetterMark%
School of Electrical Engineering and Computer Science, Washington State University, Pullman, WA, USA\\
\IEEEauthorrefmark{2}%
Mathematics and Computer Science Division, Argonne National Laboratory, Lemont, IL, USA\\
\IEEEauthorrefmark{3}%
Computational Research Division, Lawrence Berkeley National Lab, Berkeley, CA, USA
}

\maketitle
\pagestyle{plain}

\begin{abstract}
Lossy compression is one of the most efficient solutions to reduce storage overhead and improve I/O performance for HPC applications.
However, existing parallel I/O libraries cannot fully utilize lossy compression to accelerate parallel write due to the lack of deep understanding on compression-write performance.
To this end, we propose to deeply integrate predictive lossy compression with HDF5 to significantly improve the parallel-write performance.
Specifically, we propose analytical models to predict the time of compression and parallel write before the actual compression to enable compression-write overlapping.
\revise{
We also introduce an extra space in the process to handle possible data overflows resulting from prediction uncertainty in compression ratios.
}
Moreover, we propose an optimization to reorder the compression tasks to increase the overlapping efficiency.
Experiments with up to 4,096 cores from Summit show that our solution improves the write performance by up to 4.5$\times$ and 2.9$\times$ over the non-compression and lossy compression solutions, respectively, with only 1.5\% storage overhead (compared to original data) on two real-world HPC applications. 
\end{abstract}

\setlength{\textfloatsep}{6pt}
\setlength{\abovecaptionskip}{3pt}
\setlength{\abovedisplayskip}{2pt}
\setlength{\belowdisplayskip}{2pt}
\setlength{\abovedisplayshortskip}{2pt}
\setlength{\belowdisplayshortskip}{2pt}

\section{Introduction}

Large-scale scientific simulations on HPC systems play an important role in today's science and engineering domains.
Such simulations can generate extremely large amounts of data that are highly compute and storage intensive.
For example, one Nyx~\cite{almgren2013nyx} cosmological simulation with a resolution of $4096\times 4096 \times 4096$ cells can generate up to 2.8 TB of data for a single snapshot; a total of 2.8 PB of disk storage is needed, assuming the simulation runs 5 times with 200 snapshots dumped per simulation.
\revise{
Nowadays, the ever-increasing computation power can be utilized to run the simulations.
However, managing such large amounts of data remains a major challenge.
}
It is impractical to save all the generated raw data to disk due to: (1) the limited storage capacity even for supercomputers, and (2) the I/O bandwidth required to save these data can create bottlenecks in the transmission~\cite{wan2017comprehensive,wan2017analysis,cappello2019use}.

Lossy compression has been identified as one of the major data reduction techniques to address this issue.
Specifically, a new generation of error-bounded lossy compression techniques, such as SZ~\cite{tao2017significantly, di2016fast, liangerror}, ZFP~\cite{zfp}, and MGARD \cite{ainsworth2018multilevel}, have been widely used in the scientific community ~\cite{di2016fast,tao2017significantly,zfp,liangerror,lu2018understanding,luo2019identifying,tao2018optimizing,cappello2019use,jin2020understanding,grosset2020foresight}. 
Compared to lossless compression that typically achieves only $2\times$ of compression ratio~\cite{son2014data} on scientific data, error-bounded lossy compressors provide much higher compression ratio with controllable loss of accuracy.

Scientific applications running on 
Supercomputers typically use parallel I/O libraries such as 
Hierarchical Data Format 5 (HDF5)~\cite{hdf5} for managing their data. In specific, HDF5 is considered to provide high parallel I/O performance, portability of data, and rich APIs for managing data on these systems. It has been heavily used at supercomputing facilities for storing, reading, and querying scientific datasets~\cite{folk2011overview,nyx}.
This is because HDF5 has specific designs and performance optimizations for popular parallel file systems such as Lustre~\cite{byna2020exahdf5,pokhrel2018parallel}.
In addition, instead of using a general database in distributed storage, these datasets have their specific data management approach based on the parallel file system~\cite{byna2020exahdf5,pokhrel2018parallel}.
Moreover, HDF5 also provides users dynamically loaded filters \cite{hdf5filter} such as lossless and lossy compression \cite{hdf5filter-sz}, which can automatically store and access data in compressed formats.
Thus, it allows scientific applications \revise{to} store and access the data in compressed formats.
Parallel I/O in HDF5 with lossy compression filters can not only significantly reduce data size, but also improve the overall I/O performance.

\begin{figure}[]
    \centering
    \includegraphics[width=0.9\linewidth]{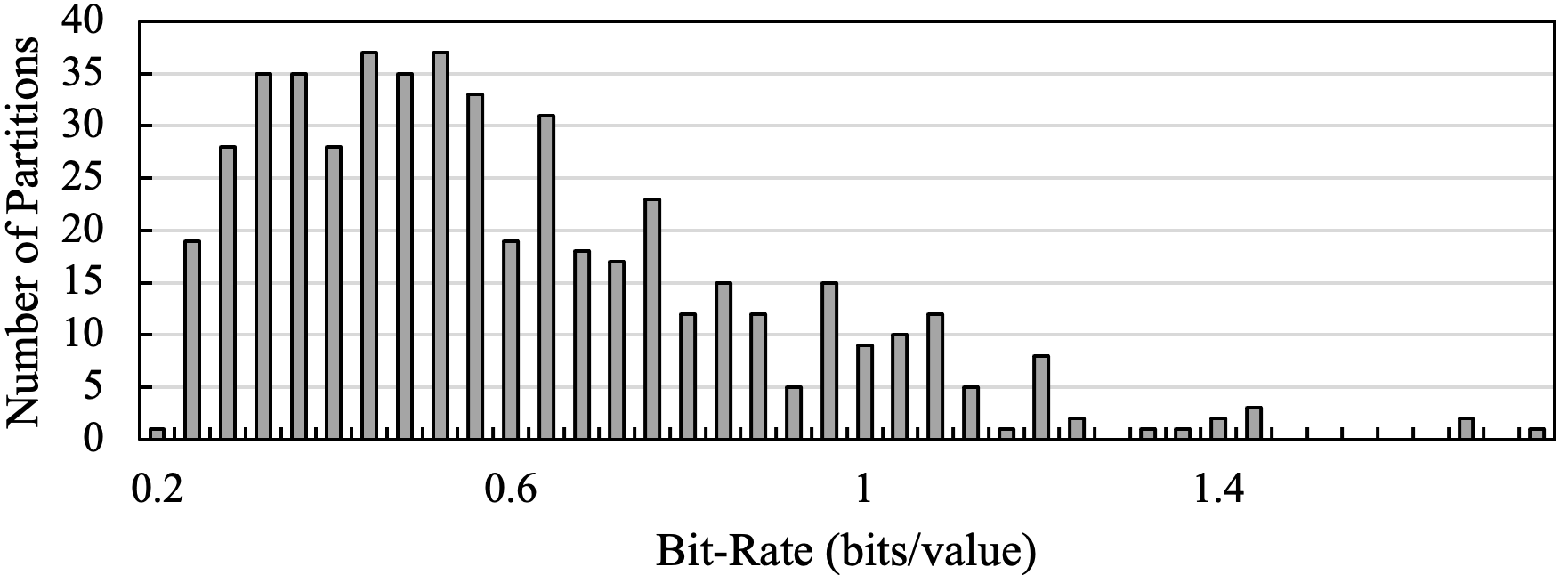}
    \caption{Compression bit-rate distribution on \revise{a Nyx dataset} with 512 partitions. Every partition uses the same compression configuration.}
    \vspace{-2mm}
    \label{fig:fig-distribution}
\end{figure}

However, the existing implementation of HDF5 with compression filters cannot fully utilize the benefits of reductions on data size and parallel-write time.
This is because to write the data from different processes to a shared file, one must specify the offset of each data partition before writing the data. Thus, the parallel write cannot start until all compression tasks finish and the compressed sizes are communicated among all processes. 
Moreover, the compressed size of each data partition is distributed across a wide range of bit-rates (i.e., bits/value) and restricts any simple pre-allocation strategy. Figure~\ref{fig:fig-distribution} shows the distribution of compressed bit-rates on \revise{a} Nyx dataset.
When using lossless compression filters, the parallel-write performance could be even lower than the original non-compression solution due to this high overhead~\cite{hdf5losless}.

To solve this issue, we propose a parallel write solution that 
integrates predictive lossy compression with the asynchronous I/O feature in HDF5, which overlaps I/O latency with compression 
to significantly improve the parallel write performance \cite{2021:TPDS:hdf5:async-io, byna2020exahdf5}.
Inspired by the previous research of Jin et al,~\cite{jin2021improving} that estimates the compression ratio of prediction-based lossy compression with little overhead, we propose to predict the time of compression and parallel write before the actual compression, and leverage the asynchronous I/O feature to overlap compression with write.  
We also introduce an extra space to handle the uncertainty of the prediction.
Moreover, we propose an optimization algorithm to reorder each process's compression tasks to increase the overlapping efficiency.
The contributions of this paper are summarized as follows.
\begin{itemize}[noitemsep, topsep=0pt, leftmargin=1.3em]
    \item We extend the prediction model for compression ratio to predict the throughputs of compression and parallel write for prediction-based lossy compression such as SZ. 
    \item We propose a new compression-write scheme with HDF5 that can efficiently write the compressed data from different processes to a shared file by overlapping compression with write based on our prediction models.
    \item We optimize the execution order of compression tasks in each process to achieve higher parallel-write performance.
    \item We evaluate our proposed solution on two real-world HPC applications with up to 4,096 cores on Summit supercomputer and up to 512 cores on Bebop cluster. Experiments demonstrate that our solution improves the parallel-write performance by up to $4.5\times$ and $2.9\times$ compared to the HDF5 write without compression and with the SZ lossy compression filter (called ``H5Z-SZ'') \cite{hdf5filter-sz}, respectively, with only $1.5\%$ storage overhead. 

    
\end{itemize}

In Section~\ref{sec:background}, we discuss the research background. In Section~\ref{sec:design}, we present our design of parallel write with compression. 
In Section~\ref{sec:evaluation}, we present our evaluation results. In Section~\ref{sec:conclusion}, we conclude our work and discuss future work.

\section{Research Background}
\label{sec:background}

In this section, we present the background information on HDF5 and lossy compression and discuss the challenges.

\subsection{Parallel I/O Libraries for HPC Applications}

\begin{figure}[]
    \centering
    \includegraphics[width=0.95\linewidth]{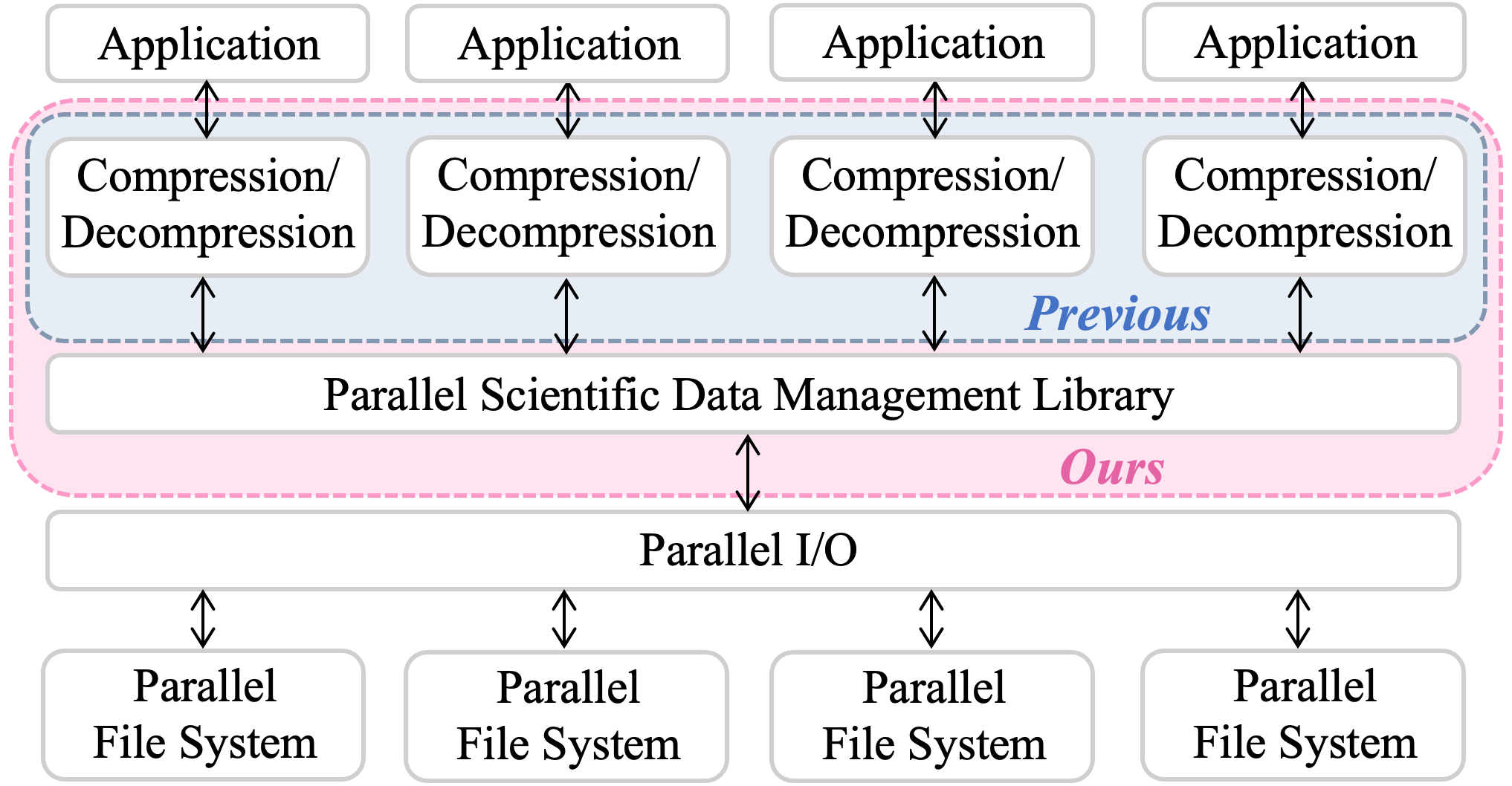}
    \caption{Scientific data management with compression.}
    \vspace{-2mm}
    \label{fig:fig-data-manage}
\end{figure}

HPC applications generate and analyze massive amounts of data. A critical requirement of these applications is the capability to access and manage this data efficiently on HPC systems. Parallel I/O becomes the key technology to enable efficiently moving data between compute nodes and storage considering the complex storage hierarchy including node-local persistent memory, burst buffers, and disk-based storage. 
For example, HDF5~\cite{hdf5}, netCDF \cite{li2003parallel}, and Adaptable IO System (ADIOS)~\cite{godoy2020adios} are the most widely used high-performance I/O libraries for HPC applications.
However, these I/O libraries still suffer from handling extremely large files (e.g. petabytes and beyond) due to inevitably limited I/O bandwidths.
Therefore, compression techniques are often adopted by them \cite{tang2021tuning}. For instance, H5Z-SZ~\cite{hdf5filter-sz} is a data filter for integrating SZ compression into HDF5.

Figure~\ref{fig:fig-data-manage} shows the abstraction of different layers in the I/O systems, where compression is an individual layer in these systems. 
Specifically, compression is normally performed between generating and storing the data 
\revise{
(e.g., H5Z-SZ).
}
It is worth noting that the compression tasks of all processes and the parallel writing of compressed data to a shared file must be performed sequentially; in other words, there must be a synchronization between these two steps.
This is because parallel writing of data from different processes to a shared file requires the size/offset of each data partition, but different processes may have vastly different data sizes after compression (even with the same size before compression).
Thus, the current compression-write scheme cannot fully utilize the high compression ratio provided by lossy compression, especially for large-scale HPC applications with multiple data fields.

In this paper, considering HDF5 
is well received by the HPC community as a system supporting parallel I/O, we mainly focus our performance evaluation on HDF5 without loss of generality.
\revise{
In addition, to improve the performance and productivity, a recent release of HDF5~\cite{byna2020exahdf5} implements virtual object layer (VOL) which can redirect I/O operations into VOL connector and allow asynchronous I/O~\cite{2021:TPDS:hdf5:async-io}. This feature enables an application to overlap I/O with other operations such as compression. 
}
Therefore, we can leverage this capability to deeply integrate and overlap predictive lossy compression with parallel write to improve the parallel write performance.
Moreover, we focus on the parallel write to a large shared file due to three main factors: 
(1) it reduces scientists' workload to manage multiple files for storage, post-hoc analysis, and visualization;
(2) it reduces the performance overhead of opening/closing multiple files and the storage overhead of metadata for many small files; 
and (3) partial processes (e.g., up to 4096 processes in \cite{byna2021tuning}) of a large-scale simulation with subfiling still writes to a shared file.






\subsection{Error-Bounded Lossy Compression}

Lossy compression can compress data with extremely high compression ratio by losing non-critical information in the reconstructed data.
Two types of most important metrics to evaluate the performance of lossy compression are: (1) compression ratio, i.e., the ratio between original data size and compressed data size, or bit-rate, i.e., the number of bits on average for each data point on average (e.g., 32/64 for single/double-precision floating-point data before compression); and (2) data distortion metrics such as peak signal-to-noise ratio (PSNR) to measure the reconstructed data quality compared to the original data.
In recent years, a new generation of high accuracy lossy compressors for scientific data have been proposed and developed for scientific floating-point data, such as SZ~\cite{di2016fast, tao2017significantly, liangerror} and ZFP~\cite{zfp}. These lossy compressors provide parameters that allow users to control the loss of information due to lossy compression \revise{precisely}. 
Unlike traditional lossy compressors such as JPEG~\cite{wallace1992jpeg} 
which are designed for images (in integers), SZ and ZFP are designed to compress floating-point data and can provide a strict error-controlling scheme based on user's requirements.
Generally, lossy compressors provide multiple compression modes, such as error-bounding mode. 
Error-bounding mode requires users to set an error type, such as point-wise absolute error bound and point-wise relative error bound, and a bound value (i.e., $10^{-3}$). The compressor ensures that 
differences between 
original
and 
reconstructed data do not exceed the 
error bound.

SZ is a prediction-based error-bounded lossy compressor for scientific data. It has three main steps: (1) predict each data point's value based on its neighboring points by using an adaptive, best-fit prediction method; (2) quantize the difference between the real value and predicted value based on the user-set error bound; and (3) apply a customized Huffman coding and lossless compression to achieve a high ratio.

\revise{
Prior works have studied the impact of lossy compression on reconstructed data quality and post-hoc analysis, providing guidelines on how to set the compression configurations for certain applications~\cite{jin2020understanding,jin2020adaptive,sz3,sz18,sz17,sz16}.
For example, a comprehensive framework was established to perform a systematic analysis on compression configurations with a given dataset and provides the best-fit solution that satisfies the post-hoc analysis~\cite{grosset2020foresight}.
Moreover, Jin et al.~\cite{jin2021improving} proposed a theoretical ratio-quality model to efficiently maximize the compression ratio given the quality constraints of post-hoc analysis.
Note that similar to the previous work \cite{zhou2021designing} on improving communication efficiency via lossy compression, this work assumes that the compression-configuration is set up by users based on their requirements of data quality, thus, the above compression-configuration methods are orthogonal to our solution. 
}

\subsection{Target I/O-Intensive HPC Applications}

In this paper, we focus mainly on two I/O-intensive HPC applications---Nyx \cite{nyx} and VPIC \cite{bowers2008ultrahigh}, which have been used for many previous I/O studies \cite{byna2012parallel, behzad2013taming, behzad2014improving, bhimji2016accelerating, han2017accelerating, koo2021empirical}. 

Nyx is an adaptive mesh, hydrodynamics code designed to model astrophysical reacting flows on HPC systems~\cite{almgren2013nyx,nyx}. This code models dark matter as discrete particles moving under the influence of gravity. The fluid in gas-dynamics is modeled using a finite-volume methodology on an adaptive set of 3-D Eulerian grids/meshes. The mesh structure is used to evolve both the fluid quantities and the particles via a particle-mesh method. For parallelization, Nyx uses MPI for the long-range force calculation and architecture-specific programming language for the short-range force algorithms, such as OpenMP and CUDA. Nyx data uses multiple 3-D arrays to represent field information in grid structure.  According to prior studies~\cite{nyx,habib2016hacc}, it can run up to millions of cores on leadership supercomputers such as Summit~\cite{summit}. 

VPIC (vector particle-in-cell) is a large-scale plasma physics simulation that can produce an unprecedented amount of data~\cite{bowers2008ultrahigh}.
Collisionless magnetic reconnection is an important mechanism that releases explosive amounts of energy as field lines break and reconnect in plasmas.
This reconnection also plays an important role in a variety of astrophysical applications that involve both hydrogen and electron-positron plasmas, including when the Earth’s magnetosphere reacts to solar eruptions which can interfere satellite communication.
Simulation of magnetic reconnection with VPIC is inherently a multi-scale problem that initiated in the small scale around individual electrons but eventually leads to a large-scale reconfiguration of the magnetic field. 
Recent simulations have revealed that electron kinetic physics is important not only in triggering reconnection, but also in its subsequent evolution~\cite{daughton2006fully}. 
This means plasma physics scientists find that they need to model the detailed electron motion, and that modeling poses severe computational challenges for 3D simulations of reconnection. 
A full-resolution magnetosphere simulation is an exascale-class computing problem.



\section{Proposed Design Methodology}
\label{sec:design}

In this section, we present our proposed design that deeply integrates the prediction-based lossy compression into HDF5 to significantly improve the parallel-write performance. 

\subsection{Design Overview}

\begin{figure}[]
    \centering
    \includegraphics[width=0.9\linewidth]{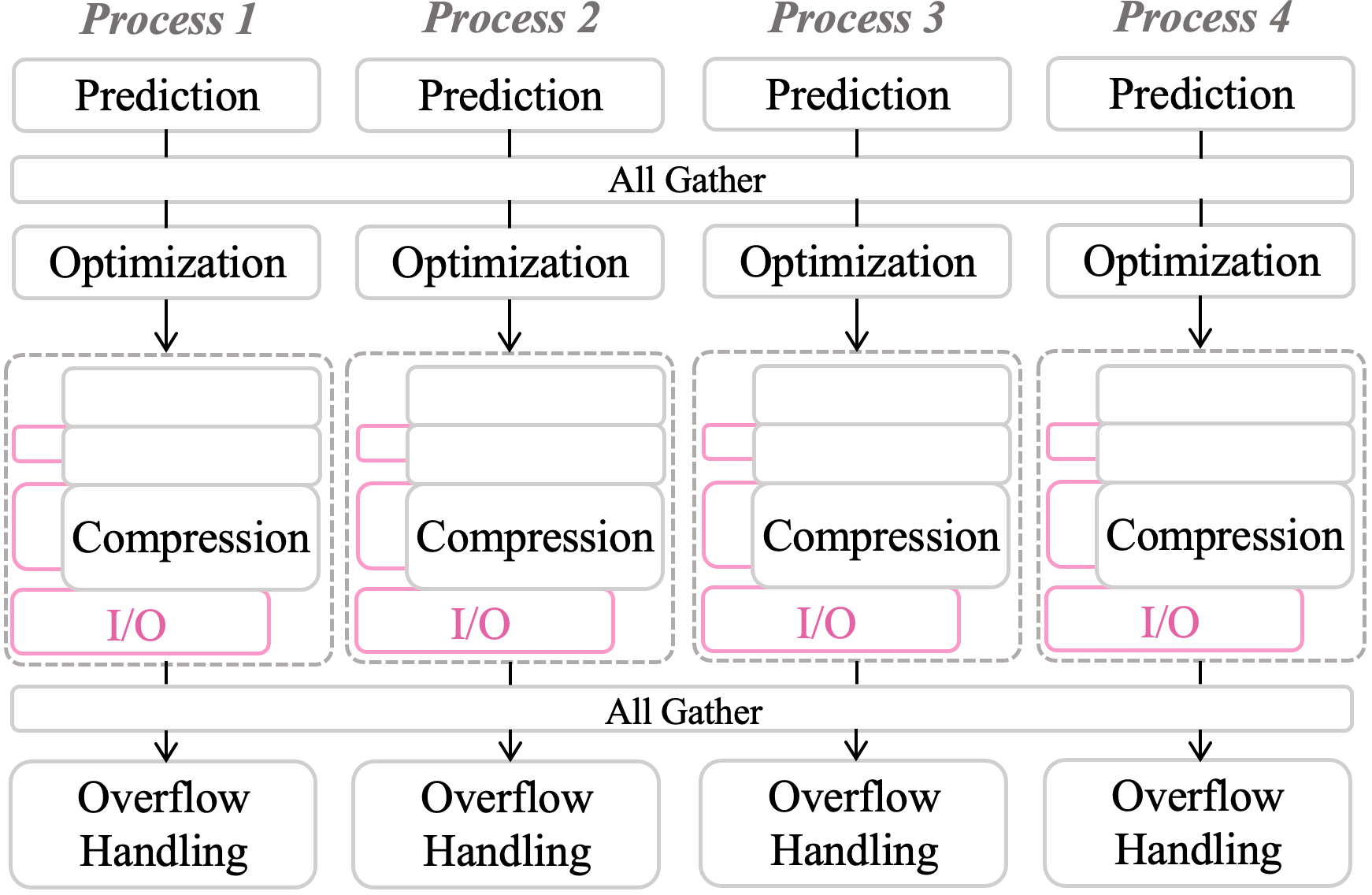}
    \caption{Overview of our proposed solution.}
    \vspace{-2mm}
    \label{fig:fig-overview}
\end{figure}

\begin{figure*}[]
    \centering
    \includegraphics[width=.95\linewidth]{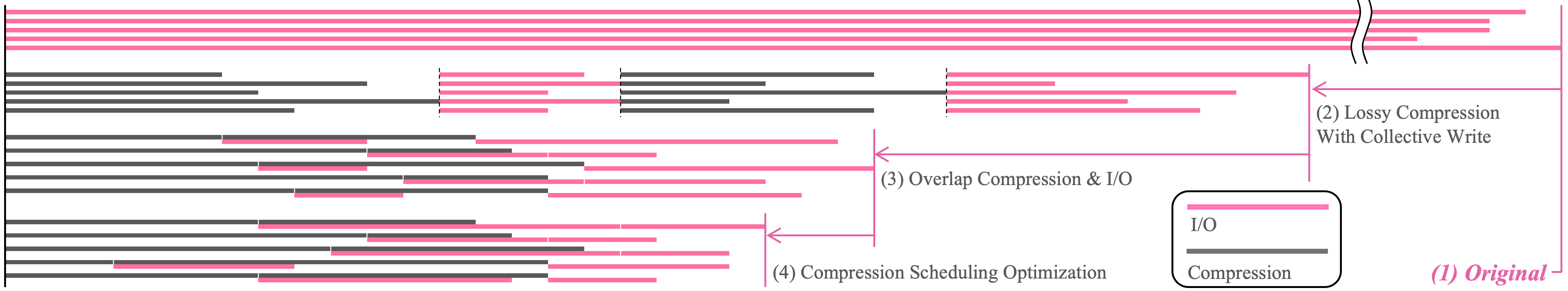}
    \caption{Timeline of data aggregation with 5 processes and 2 data fields.}
    \vspace{-4mm}
    \label{fig:fig-timeline}
\end{figure*}

Figure~\ref{fig:fig-overview} shows an overview of our proposed framework. 
Specifically, we first conduct the ratio and throughput prediction phase for all data partitions on each process, which includes estimations of compression ratio, compression throughput, and I/O throughput. 
Then, an all-gather communication is performed to distribute the estimated compression ratio of each partition to all processes. 
Note that the estimations of compression throughput and I/O throughput are not distributed in this phase, since each process only requires these estimations of its own data partition in order to perform the following optimization.
Next, each process computes the offset of its own data partition of each field for parallel write based on the estimated compressed data size 
with an extra space (a mechanism to handle overflow data under unexpected prediction failure).
Note that since each process gathers the same estimation results from others, the computation of offsets is consistent across all processes.
After that, we perform an optimization on the order of compressing different data fields in each process. 
Lastly, we overlap compressions and writes based on the predicted write offsets and the optimized compression order (the order of compressing different fields).

Compared to the non-compression solution that writes the entire data to a shared file, our solution can significantly improve the overall write performance due to the high compression ratio of lossy compression that reduces I/O traffic.
Compared to the previous lossy compression solution using HDF5 filter, our solution can overlap compression and write that allows independent and asynchronous write across processes. 
Figure~\ref{fig:fig-timeline} shows the simplified timeline of parallel write from five processes to a shared file on two data fields with four methods: 
\underline{(1)} collective write without compression; 
\underline{(2)} collective write with compression via filter; 
\underline{(3)} independent writes overlapped with compressions; and 
\underline{(4)} independent writes overlapped with reordered compressions. 

Specifically, the improvement from method \underline{(2)} to method \underline{(3)} is due to the overlapping.
From method \underline{(3)} to method \underline{(4)}, the scheduling optimizer reordered some compression tasks in each process to further improve the overlapping efficiency.
For example, the optimizer switches the compression order of the two data fields in the last process, i.e., the data with smaller compressed size are compressed later. 
The improvement of this optimization is more significant when more data fields are compressed in parallel write as it has higher chance to improve the overlapping efficiency. We will discuss more observations and insights in detail in Section~\ref{sec:eval-overall}.

In this paper, we focus on accelerating parallel write due to two main reasons: (1) HPC simulations are mostly write-oriented \cite{paul2020understanding}, and (2) parallel I/O libraries usually have lower performance in write than in read \cite{li2003parallel, lee2022case, byna2020exahdf5, lang2009performance, hdf5Tuning}.

In the following sections, we first introduce the prediction models for prediction-based lossy compressors such as SZ.
Next, we present the estimation of I/O throughput based on our models and empirical studies. 
After that, we propose our compression-write overlapping design with HDF5.
Finally, we propose our algorithm for compression scheduling.

\subsection{Compressor Throughput Estimation}

The estimation on compression ratio in this work is based on recent research~\cite{jin2021improving}, which proposed a ratio-quality model \revise{(i.e., compression ratio and reconstructed data quality model)} for prediction-based lossy compressor. 
It developed algorithms to predict the compression ratio based on a newly designed sampling strategy without performing compression, significantly reducing the time overhead of the compression ratio prediction to less than $10\%$ (in relative to the compression time). 
We leverage this idea to enable the overlapping design with small prediction overhead in our design.

On the other hand, estimating the compression throughput is also essential for compression order optimization.
Generally, higher compression ratio results in higher compression throughput~\cite{jin2020understanding}, due to the shorter time for building Huffman tree and encoding with smaller tree.
However, there is no prior work for an accurate compression-throughput estimation.

\begin{figure}[]
    \centering
    \includegraphics[width=0.9\linewidth]{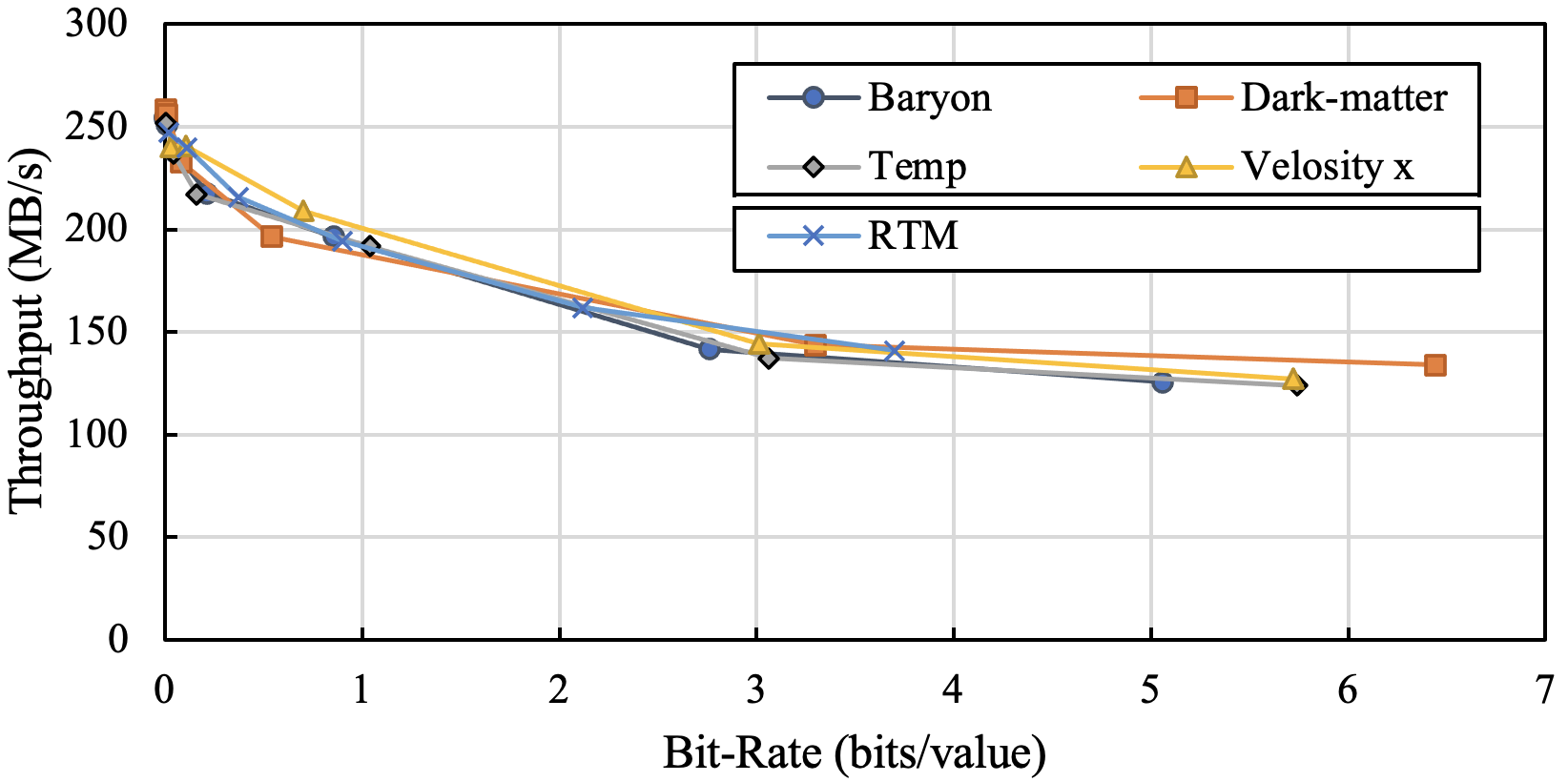}
    \caption{Single-core compression throughput with different bit-rates on a Nyx and a RTM datasets. Evaluated on Bebop.}
    \label{fig:fig-comp-throughput}
\end{figure}

In this work, we propose to estimate compression throughput based on the predicted compression ratio for each data partition.
Figure~\ref{fig:fig-comp-throughput} shows our empirical evaluation of compression throughput and compression ratio with multiple data fields and data types 
across different error bounds.
Note that we use bit-rate (the average bits used to represent each value) on the x-axis to better illustrate the size of compressed data.
For example, a bit-rate of 4 is equivalent to a compression ratio of $8\times$ when compressing single-precision floating-point data (32 bits/value).
The evaluation is performed on a single CPU core with SZ lossy compressor, since our target method (4) is to compress the data in each process/core independently before writing the compressed data to the shared file.
The figure illustrates that (1) the maximum and minimum compression throughput are similarly bounded across different data samples (i.e., about 120$\sim$250 MB/s); and (2) 
the bitrate-throughput curve for each data sample is highly consistent.

\begin{figure}[]
    \centering
    \includegraphics[width=0.85\linewidth]{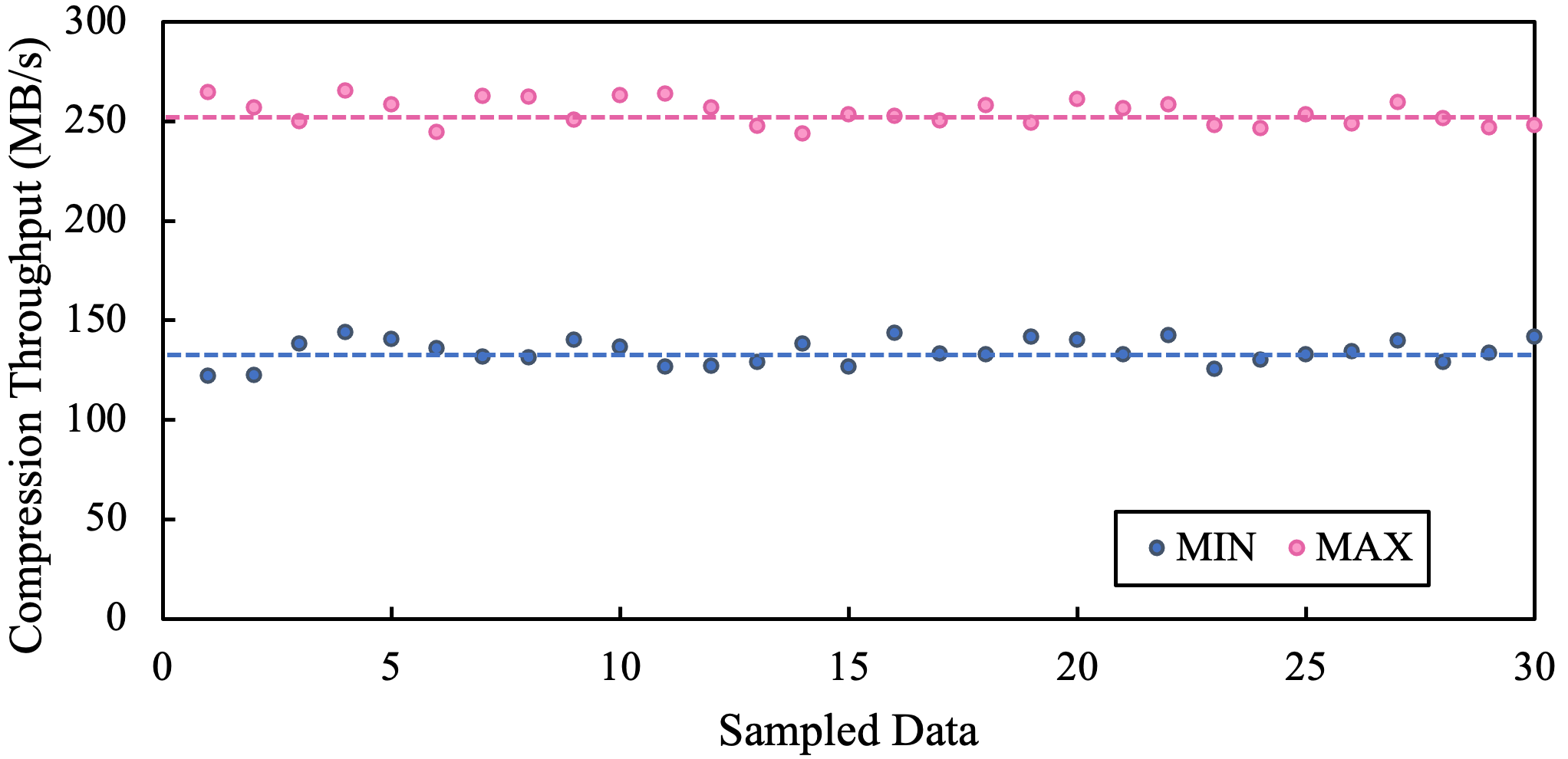}
    \caption{Minimum and maximum compression throughput of a given data partition based on 30 samples from Baryon density, dark matter density, temperature and velocity x data fields in a Nyx dataset.}
    \vspace{-2mm}
    \label{fig:fig-minmax}
\end{figure}

The reasons behind the first observation are: (1) For extremely high error bounds, despite the small size of the Huffman tree (negligible tree building time), the prediction and encoding process still passes each point, which provides an upper bound on the throughput; (2) For extremely low error bounds, since SZ limits the maximum size of the Huffman tree (the maximum number of quantization codes), if one point is unpredictable with the maximum number of quantization codes, its original value will be saved directly, which provides a lower bound on the throughput.
Figure~\ref{fig:fig-minmax} shows the maximum and minimum compression throughput evaluated on multiple data samples size of $67.1$ MB from different data fields and types on the same experiment platform.

The reason behind the second observation is that for different data samples, the quantization codes after prediction are all centrally distributed,
and the overall bit-rates are similar under the same Huffman encoding efficiency, where the sizes of the Huffman trees are also similar, resulting in stable tree building time and encoding time. 

Based on Figure~\ref{fig:fig-comp-throughput}, we use a power function to predict compression throughput
\begin{align}
 \small\textstyle
    T_{comp} &= D/S \label{equ-1}  \\
    &=(B_{ori} \times n)/(((C_{max}-C_{min})\times 3^{-a})B^{a} + C_{min}), \notag
\end{align}
where $T_{comp}$ is the estimated compression time; $D$ is the size of the original data; $S$ is the compression throughput; \revise{$B$ is the compressed bit-rate (i.e., the average bits used to save each value, e.g., a compression ratio of 16 from 32 bits floating point values results in $B=2$);} $B_{ori}$ is the original bit-rate (i.e., 32 for single-precision floating point values); $n$ is the number of points in the current data partition; $C_{min}$ and $C_{max}$ are the minimum and maximum compression throughputs, respectively; $a$ is a hyper-parameter to describe the shape of the power function \revise{(i.e., a value smaller than 0; 
the lower the value, the more curved the function).}
The compression-throughput estimation is a power function that takes min-max into account.
Note that the number 3 in the equation is based on our experiment that yields the best result.

This prediction model can be adapted for different machines, where we perform lossy compression on a sample of dataset offline to evaluate the minimum and maximum compression throughput and map it based on Equation~(\ref{equ-1}). We show the accuracy of our proposed model in Section~\ref{sec:eval-accuracy}.

\subsection{Write Time Estimation}

The write time estimation is needed when optimizing the compression order, since the compressed data with larger estimated write time tends to be processed earlier to maximize the I/O traffic occupancy.
However, compared with the estimation on compression ratio and throughput, the accuracy required for write-time estimation is considerably lower.
This is because one inaccurate write-time prediction makes a linear shift to all other write-time predictions within this process, 
\revise{
so the relative time spent for each write remains unchanged compared to the entire time,
}
and the optimization would not be affected (but only increase or decrease the actual write time).
As a result, our goal is not to provide a highly accurate write-time estimation for each data partition, but to provide a capability to estimate the relative write time 
across different data sizes.

\begin{figure}[]
    \centering
    \includegraphics[width=0.86\linewidth]{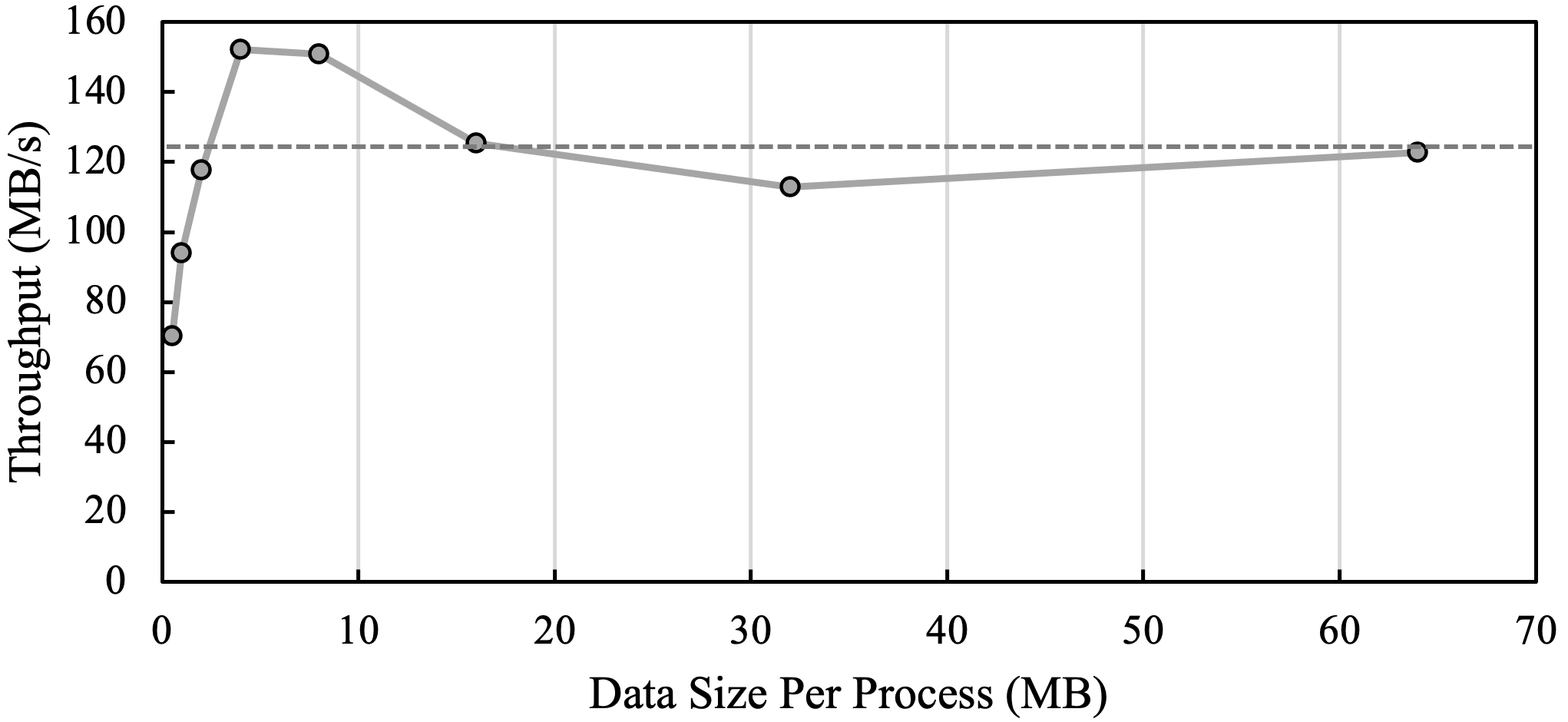}
    \caption{Independent write I/O throughput per process with different data sizes per process. Evaluated on 128 processes.}
    \label{fig:fig-IO-thr}
\end{figure}

Figure~\ref{fig:fig-IO-thr} shows the parallel-write throughput per process with different data partition sizes.
We can observe that 
the average throughput first increases as the data size increases and stabilizes after the data size reaches a certain point.
Similar to the compression-throughput estimation, the write-time estimation is also based on the predicted bit-rate.
Note that the compression throughput is based on the uncompressed data size. Considering that usually each process handles a similar amount of data during the simulation, the variance of the compression-time difference is limited to $T_{max}/T_{min}$.
While the write throughput is based on the compressed data size, which can easily vary over $10\times$ across data partitions according to our evaluation.
This means the write time is mainly dependent on the compression ratio assuming the write throughput is relatively stable across different processes without I/O congestion.
Thus, we estimate the write time as
\begin{align}
 \small\textstyle
    T_{write} = (B\times n)/C_{thr},
    \label{equ-2} 
\end{align}
where $T_{write}$ is the write time, $B$ is the compressed bit-rate; $n$ is the number of points in the data partition; and $C_{thr}$ is the stable write throughput based on an empirical evaluation.

\subsection{Overlapping Compression and Write}
\label{sec:3-redundant}

\begin{figure}[]
    \centering
    \vspace{-4mm}
    \includegraphics[width=0.95\linewidth]{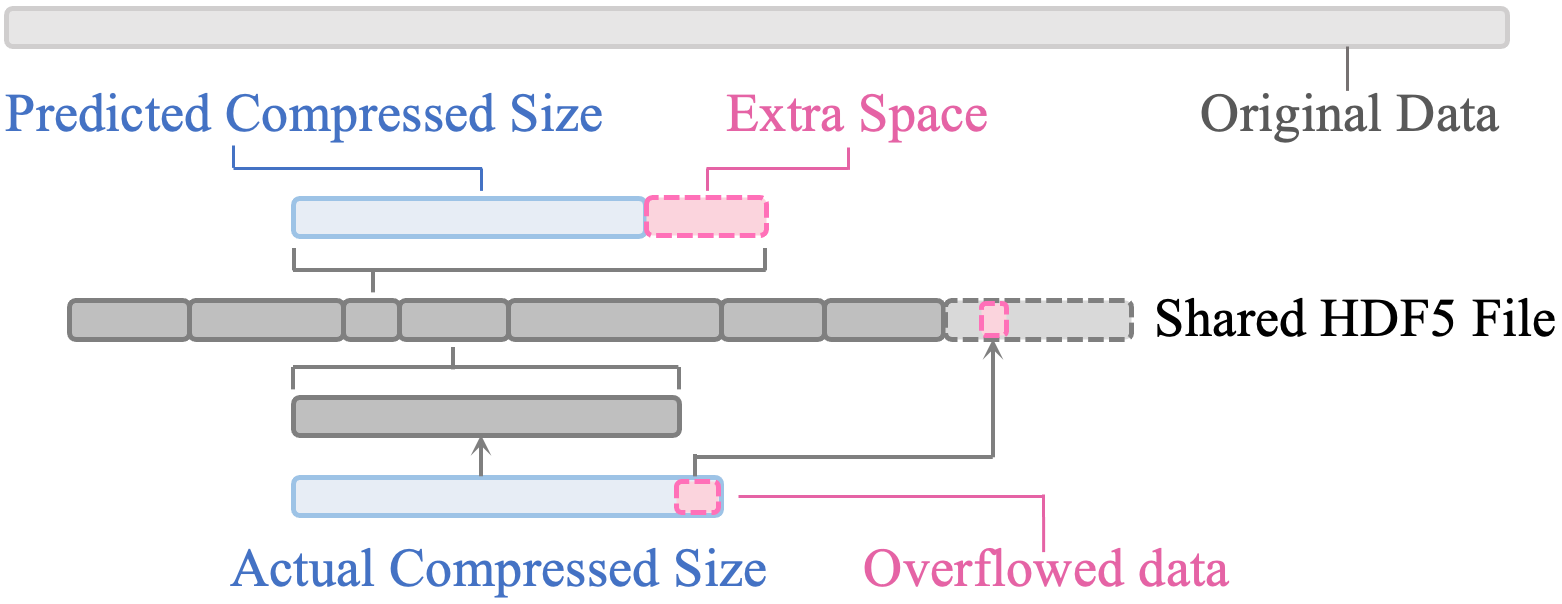}
    \caption{Overflow data handling with preserved extra space.}
    \vspace{-2mm}
    \label{fig:fig-redundant}
\end{figure}

When writing data from different processes to a shared file, the well-known I/O libraries such as HDF5, MPI-IO, and ADIOS require users to provide an offset for each data partition. 
For the original non-compression parallel write, the offset of each partition can easily be computed based on the data size.
However, with compression, the size of each compressed partition can vary drastically due to diverse data compressibility and different compression configurations.
Prior to our work, it was impossible to pre-compute the offset before the compression, thus, it must sequentially process the collective write after compression. 
Thanks to the compression-ratio estimation for each partition~\cite{jin2021improving}, we can estimate the offset even before actual compression and allow overlapping between compression and write via HDF5's asynchronous I/O. 

We note that although the compression-ratio model can provide high accuracy for prediction-based lossy compression, 
the prediction does not have a guarantee bound. 
This means that we must handle the situation when the actual compressed data size is larger than the predicted size,
since it is impractical to simply recompute and update the offsets of all following partitions every time when the prediction failure happens as it would introduce a large amount of communication overhead.
To this end, we propose to reserve an particular extra space for each prediction to handle the possibility of compressed data overflow.
Moreover, we also propose an overflow data handling scheme to redirect and store the exceeded portion beyond the extra space. 
Figure~\ref{fig:fig-redundant} shows our proposed overflow handling strategy with an extra data space.

\begin{figure}[]
    \centering
    \includegraphics[width=0.95\linewidth]{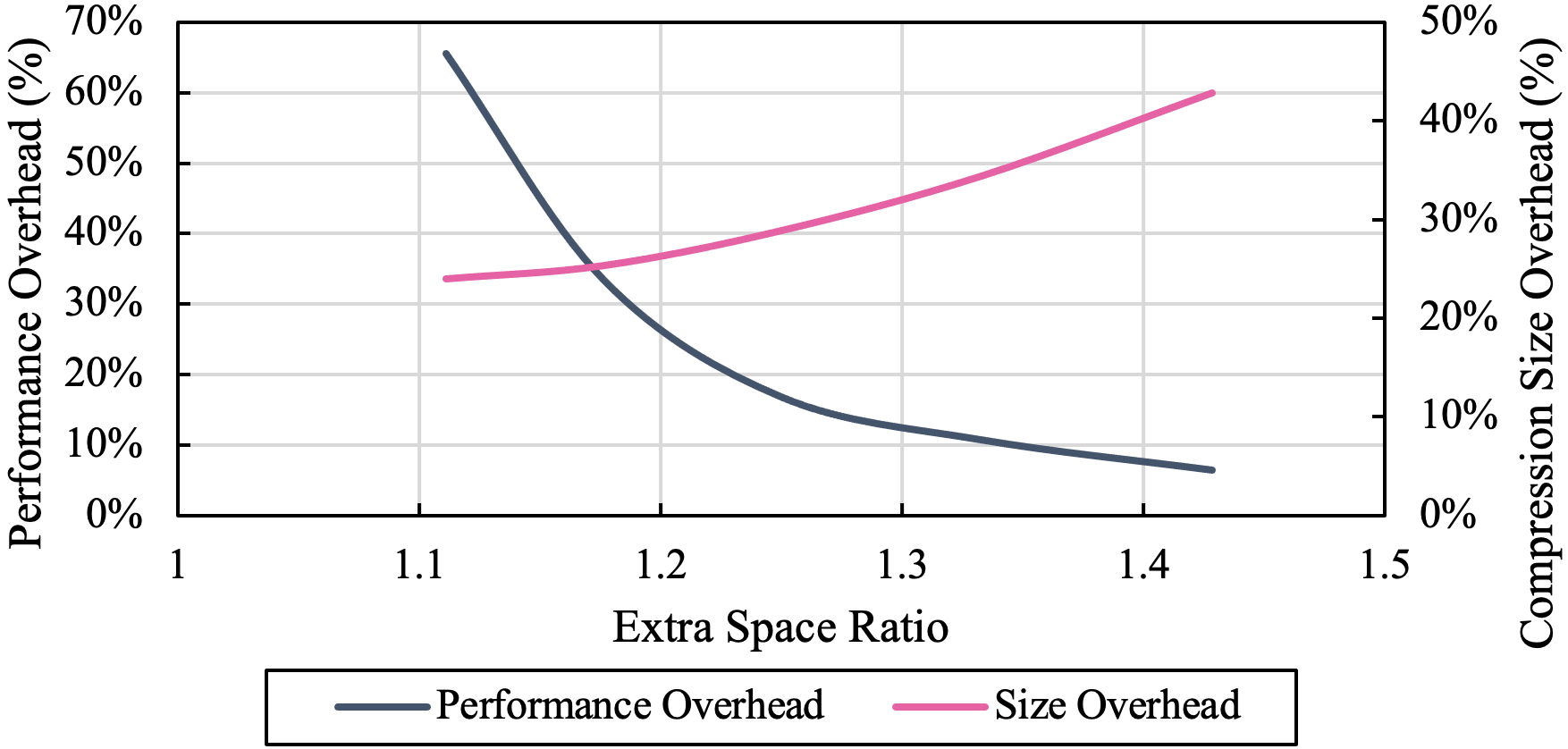}
    \caption{Trade-off between performance overhead and compression size overhead. Shown the empirical average result based on Nyx and VPIC datasets on 512 processes.}
    \label{fig:fig-redundant-2}
\end{figure}

As the compression-ratio estimation error is not bounded, we cannot simply set a maximum extra space to guarantee no data overflow in practice.
Thus, we take the extra space ratio as one of the tunable parameters in our framework.
For example, setting this ratio $R_{space}$ to 1.5 means that we over preserve 50\%  storage space (in relative to the predicted compressed size) when computing the offset of each partition.

In our evaluation, we find that the compression-ratio model~\cite{jin2021novel} performs poorly under extremely high compression ratio because of three factors: (1) when the compression ratio is high, the Huffman encoder can only provide a maximum compression ratio of $32\times$ (when the original data is single-precision) and relies on the following lossless compression to further reduce the encoded data; (2) the compression-ratio model is based on run-length encoding to analyze the lossless encoding efficiency, which naturally features lower estimation accuracy compared to the efficiency estimation of Huffman encoding; and (3) when the compression ratio is low, the encoded data stream from the Huffman encoder is highly random and is hardly further reduced by lossless compressors.
Thus, when handling the data partition with compression ratio higher than 32 (bit-rate lower than 1), we set $r_{space}$ to
\begin{align}
 \small\textstyle
    r_{space} = & \min(2, 1+(R_{space}-1)\times4), \notag \\
    & where \hspace{5mm} r_{comp} > 32.
    \label{equ-2} 
\end{align}


where $r_{comp}$ is the compression ratio.
\revise{
On one hand, the higher the extra space ratio is, the larger the storage size of the compressed shared file will be, due to the larger extra space that is likely to be wasted.
}
On the other hand, increasing $r_{space}$ may improve the overall performance due to less data overflow processing overhead, and vice versa.
Overall, above a certain threshold (e.g., $r_{space}>1.1\times$ in our evaluation), increasing $r_{space}$ becomes a trade-off between the write performance and the overall compression efficiency.

More specifically, below this threshold, decreasing $r_{space}$ results in a significant write performance drop due to a large number of compressed data overflows.
For example, when $r_{space} = 1.1\times$, $32.4\%$ data partitions suffer from data overflow, causing an extra $65.6\%$ time overhead. 
\revise{Based on our observations from Figure~\ref{fig:fig-sec4-redundant},} the trade-off between write performance and compression efficiency is relatively similar across different fields and datasets.
We will demonstrate this with more details in Section~\ref{sec:eval-redundant}.

As a result, we propose to weight the write performance and compression efficiency overhead and provide a tunable ratio between them for users.
For any given weights, we provide the extra space ratio based on our proposed mapping.
Figure~\ref{fig:fig-redundant-2} shows our mapping solution for the extra space ratio (a more detailed evaluation illustrating the accuracy of this mapping will be shown in Section~\ref{sec:eval-redundant}).
Note that we only support $r_{space}$ between $[1.1, 1.43]$ due to (1) an extremely high time overhead below 1.1, and (2) a low efficiency of trading storage for performance after 1.43.
We set the default extra space ratio to 1.25 in this work.

After estimating the offset of each data partition, we must store this offset information as metadata for the decompression purpose.
Compared to the compressed size of the entire dataset, this metadata size is totally negligible.
For example, when writing a total size of 2.5 TB Nyx dataset from 4,096 processes with SZ, there is only 295 KB metadata to store the offsets for 9 fields, while the compressed data size is 210 GB.

For handling the compressed data overflow, our goal is to append the exceeded data to the end of the shared file.
Specifically, we continue to optimize and write the maximum amount of data into the preserved area of the shared file.
Then, after all processes finish writing, we initiate an all-gather operation across all processes to distribute the size information of overflow data.
After that, some processes calculate the offset of their own excess and write it to the end of the shared file independently.
Note that since only a small fraction of processes may need to process/save a small amount of overflow data, the overall time overhead for processing overflow data is small.
A more detailed evaluation will be presented in Section~\ref{sec:eval-overall}.

\subsection{Compression Order Optimization}

\begin{algorithm}[t]
    \footnotesize
    \caption{Compression Order Optimization}
    \label{algo:algo}
    \parbox[l]{\linewidth}{\hspace*{\algorithmicindent}%
    {\bfseries Notation:} %
        data fields in current process: $\ell$; compression queue: $Q$; compression queue after insert and additional data: $Q^\circ$; possible insert locations in a queue: $\beta$; time to compress: $t_c$; time to write:$t_w$; predicted compression time: $P_c(\ell)$; predicted write time: $P_w(\ell)$
    }
\parbox[l]{\linewidth}{\hspace*{\algorithmicindent}%
    {\bfseries Global:} %
        $P_c(\ell), P_w(\ell)$
    }
\begin{algorithmic}[1]
\Procedure{Time}{$q$}
    \State \textcolor{black}{$t_c, t_w \gets 0$}
    \For{$\ell \gets$ data fields in $q$}
        \State $t_c \gets t_c + P_c(\ell)$ 
        \State \textcolor{black}{$t_w \gets P_w(\ell) + \max(t_c, t_w)$}
    \EndFor
    \State \Return $t_w$
\EndProcedure
\State

\Procedure{SchedulingOptimizator}{}
    \For{$\ell \gets$ data fields in current process}
        \For{$\beta \gets$ all possible insert location} 
        \State $Q^\circ \gets $ insert $\ell$ to $\beta$
        \If{$ \textsc{Time}(Q^\circ) < \textsc{Time}(Q)$ or first $\beta$}
            \State $Q \gets Q^\circ$
        \EndIf
        \EndFor
    \EndFor
    \State \Return $Q$
\EndProcedure
\end{algorithmic}
\end{algorithm}

As shown in Figure~\ref{fig:fig-timeline}, when writing data to the shared file, overlapping the compression and I/O can provide a significant performance benefit over the previous collective-write solution with lossy compression.
However, with multiple data fields in each process, compressing them sequentially in the original order is not the optimal solution.
In fact, since we use the modeling approach to predict the compressed size, the compression time, and the write time of each data partition, we can reorder the compression tasks for different fields to maximize the overlapping without any penalty.

Algorithm~\ref{algo:algo} shows the pseudocode for optimizing the compression order in each process.
The proposed method is based on the observation that the total compression time is theoretically fixed regardless of the compression order.
Our optimization focuses on the dependencies and timing of launching write operations for each compressed data to minimize timeouts compared to compression.
The time complexity of the proposed algorithm is $O(n^2)$,
\revise{
whereas the time complexity of our compression is $O(N)$.
Considering that $N$ (i.e., the number of values) in one data partition is significantly larger than $n$ (i.e., the number of data fields), the optimization overhead is almost negligible compared to the actual compression and write time. 
For example, this optimization overhead is only 0.17\% of the compression time under an extreme condition where $N$ is small at 32768 ($32\times32\times32$) but $n$ is very large at 100.
}

\begin{figure}[]
    \centering
    \includegraphics[width=0.95\linewidth]{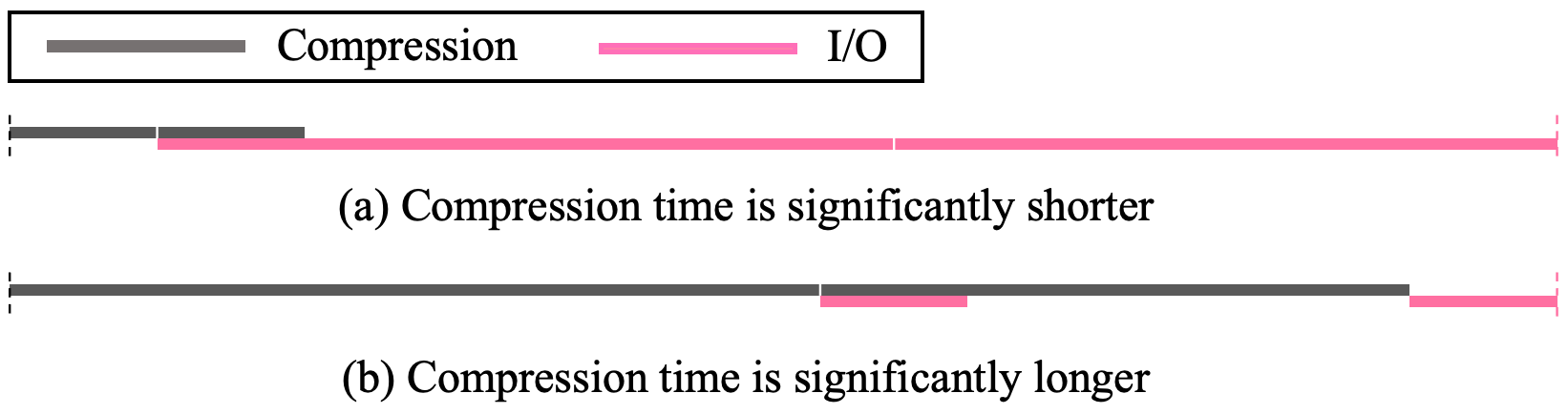}
    \caption{\revise{An example of extremely unbalanced compression time and write time, limiting the benefit from our reordering.}}
    \label{fig:fig-balance}
\end{figure}

Based on our algorithm design, we can expect the optimization to bring benefit when the balance between compression time and I/O time is relatively stable.
However, we notice that in the extreme scenarios, the compression time and the write time could be very unbalanced, which can diminish the benefit of our reordering optimization, as shown in Figure~\ref{fig:fig-balance}.
Specifically, there are two cases: 
(a) when the write time is significantly longer than the compression time, or
(b) when the compression time is significantly longer than the write time.
In both cases, there is no significant room to improve performance because of the limited overlap between compression and I/O (more details will be discussed in Section~\ref{sec:eval-overall}).

In addition, we note that when the number of data fields is relatively large, our optimization can bring greater benefit. 
This is because the overall performance is dependent on the worst process with the longest time among all the processes due to independent asynchronous writes. 


\section{Experimental Evaluation}
\label{sec:evaluation}

In this section, we present the evaluation results of our proposed framework for accelerating parallel write.
We first evaluate the accuracy of our prediction for compression time and write time. Next, we evaluate the extra space setup under different compression ratios and scales. Finally, we perform performance evaluation and scaling study of our approach and compare it with the original solution without compression and the solution using 
using the H5Z-SZ filter~\cite{hdf5filter-sz}.

\subsection{Evaluation Setup}

\begin{table}[]
\renewcommand*{\arraystretch}{1.4}
\centering
\ttfamily
\scriptsize
\caption{Details of Tested Datasets}
\newcommand\alignmiddle[2]{
\makebox[3em][r]{$(#1$}\makebox[.8em]{$,\ $}\makebox[2em][l]{$#2)$}}



\newcommand{\SUPERBOLD}{\fontfamily{ugq}\selectfont}

\resizebox{\linewidth}{!}{
\begin{tabular}{@{} l|c|c|c @{}}
\hline
\SUPERBOLD Name  & \SUPERBOLD Description    & \SUPERBOLD Scale     & \SUPERBOLD Size        \\ \hline
\multicolumn{1}{c|}{\multirow{4}{*}{nyx~\cite{nyx}}}  & \multicolumn{1}{c|}{\multirow{4}{*}{Cosmology simulation}}    & 4096$\times$4096$\times$4096     & 2.47 TB   \\ \cline{3-4}
\multicolumn{1}{c|}{}   & \multicolumn{1}{c|}{}    & 2048$\times$2048$\times$2048     & 206.15 GB   \\ \cline{3-4}
\multicolumn{1}{c|}{}   & \multicolumn{1}{c|}{}   & 1024$\times$1024$\times$1024     & 25.76 GB   \\ \cline{3-4}
\multicolumn{1}{c|}{}   & \multicolumn{1}{c|}{}   & 512$\times$512$\times$512     & 3.22 GB   \\ \hline
VPIC~\cite{byna2013trillion}    & Particle simulation   & 161,297,451,573     & 4.62 TB  \\ \hline
\end{tabular}
}

\label{tab:DataDetail}
\end{table}

We rigorously implement our approach with HDF5~\cite{byna2020exahdf5} and SZ3 \cite{sz3} (a modularized prediction-based lossy compressor).
We conduct our experiments on two HPC systems: (1) the Summit supercomputer~\cite{summit} at Oak Ridge National Laboratory, each node of which is equipped with two IBM POWER9 processors with 42 physical cores and 512 GB DDR4 memory, and (2) the Bebop cluster \cite{BebopLab56:online} at Argonne National Laboratory, each node of which is equipped with two 18-core Intel Xeon E5-2695v4 CPUs and 128 GB DDR4 memory.

We use different scales of Nyx and VPIC datasets in our evaluation.
Table~\ref{tab:DataDetail} shows the details of our tested datasets.
According to the previous work~\cite{jin2020understanding, jin2020adaptive}, using the absolute error bounds of $(0.2, 0.4, 1e+3, 2e+5, 2e+5, 2e+5)$ for compressing the six Nyx data fields (i.e., baryon density, dark matter density, temperature, velocity x, velocity y, velocity z) can satisfy the post-hoc analysis quality
\revise{
with an average PSNR (peak signal-to-noise ratio) at 78.6 dB, resulting in a compression ratio of $\sim$$16\times$.
}
The $4096\times4096\times4096$ Nyx dataset has three additional fields: particle\_vx, particle\_vy and particle\_vz.
\revise{
Similarly, we compress them with a compression ratio of $16\times$ to satisfy the post-hoc analysis quality.}
For the VPIC dataset, we also compress the 8 data fields with a compression ratio of $13.8\times$, which is suggested by the application developers according to their post-hoc analysis. 
\revise{
Note that our solution can also be applied to other scientific datasets with similar performance improvement expected.
This is because: 
(1) our solution provides the same reconstructed data quality compared to previous solutions, where many studies ~\cite{jin2020understanding,jin2020adaptive,sz3,sz18,sz17,sz16,grosset2020foresight} show that lossy compression has a high data reduction capability on a variety of applications;
(2) a previous study shows that the accuracy of the compression-ratio estimation used in our solution is consistently above 90\% across tens of benchmark datasets~\cite{jin2021improving}, implying that the proportion of data overflows due to inaccurately estimated compression ratios is relatively stable in most scientific datasets;
(3) the accuracy of the compression-throughput estimation is consistently high across different datasets and fields, as shown in Figure~\ref{fig:fig-comp-throughput}, \ref{fig:fig-accuracy} and \ref{fig:fig-accuracy-2};
and (4) the accuracy of I/O-throughput estimation only relies on the estimated compressed data size as discussed in (2).
}


\subsection{Accuracy of Compression and I/O Throughput Estimation}
\label{sec:eval-accuracy}

\begin{figure}[]
    \centering
    \includegraphics[width=\linewidth]{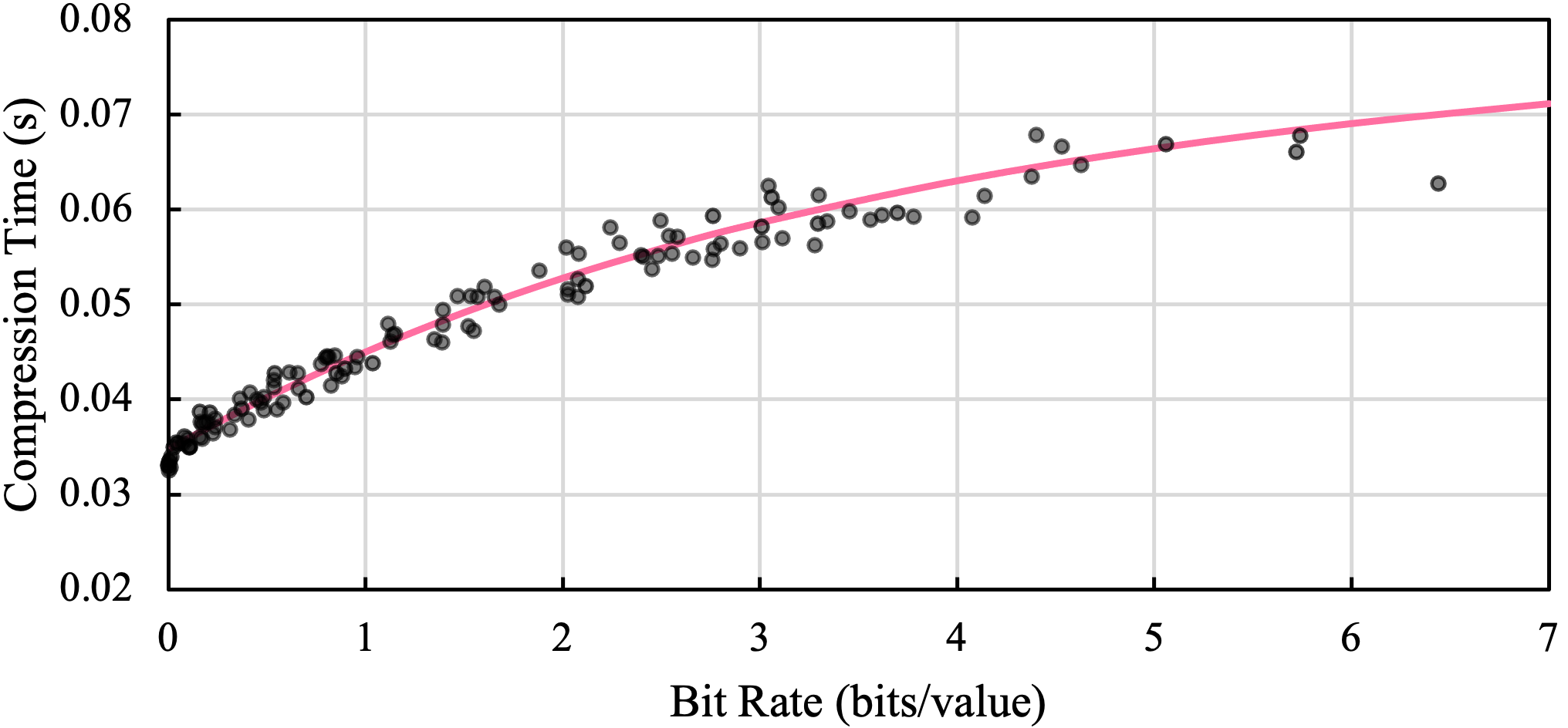}
    \caption{Accuracy of our compression-time estimation on $512^3$ Nyx data samples (red line is predicted time; black dots are actual time).}
    \vspace{-2mm}
    \label{fig:fig-accuracy}
\end{figure}

\begin{figure}[]
    \centering
    \includegraphics[width=\linewidth]{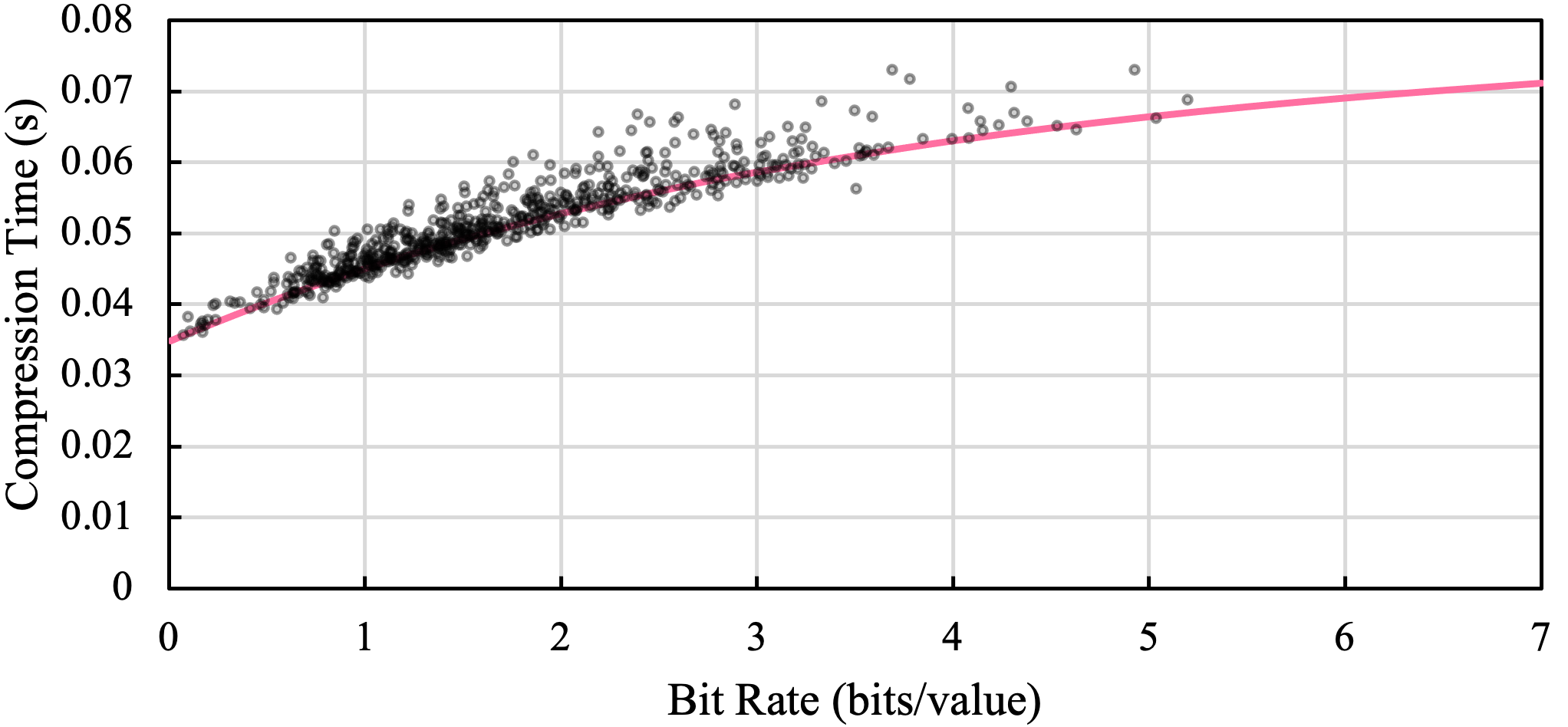}
    \caption{Accuracy of our compression-time estimation on $1024^3$ Nyx data samples. Red line is predicted time; black dots are actual time. Offline parameter is based on the baryon density of $512^3$ Nyx dataset.
    }
    \label{fig:fig-accuracy-2}
\end{figure}

First, we evaluate the accuracy of our proposed estimation on compression time. 
We perform the evaluation on the $512^3$ Nyx dataset only using the baryon density field and conduct the offline compression with the relative error bound ranging between [1e-1, 1e-8].
Then we calculate $C_{min}$, $C_{max}$ and $a$ in Equation~(\ref{equ-1}) based on the offline results, so that we can use Equation~(\ref{equ-1}) to predict the compression time of any given data (i.e., $101.7$, $240.6$, and $-1.716$, respectively, in this case).
Next, we distribute the Nyx dataset to 64 processes (each process has a $128\times128\times128$ data partition) and perform the write operation. 
For each field of data partition in each process, we predict the compression time using Equation~(\ref{equ-1}).
Figure~\ref{fig:fig-accuracy} shows the actual compression time versus the predicted compression time. 

Furthermore, we extend our evaluation to the $1024^3$ Nyx dataset and 512 processes using the same $C_{min}$, $C_{max}$, and $a$ obtained from the $512^3$ dataset.
Figure~\ref{fig:fig-accuracy-2} shows the actual compression time versus the predicted compression time. We can observe from both figures that our compression-time estimation has high accuracy even though the offline experiment only uses baryon density as input.
This is consistent with the result shown in Figure~\ref{fig:fig-comp-throughput}, where different data fields and datasets have similar compression throughputs.

\begin{figure}[]
    \centering
    \includegraphics[width=1\linewidth]{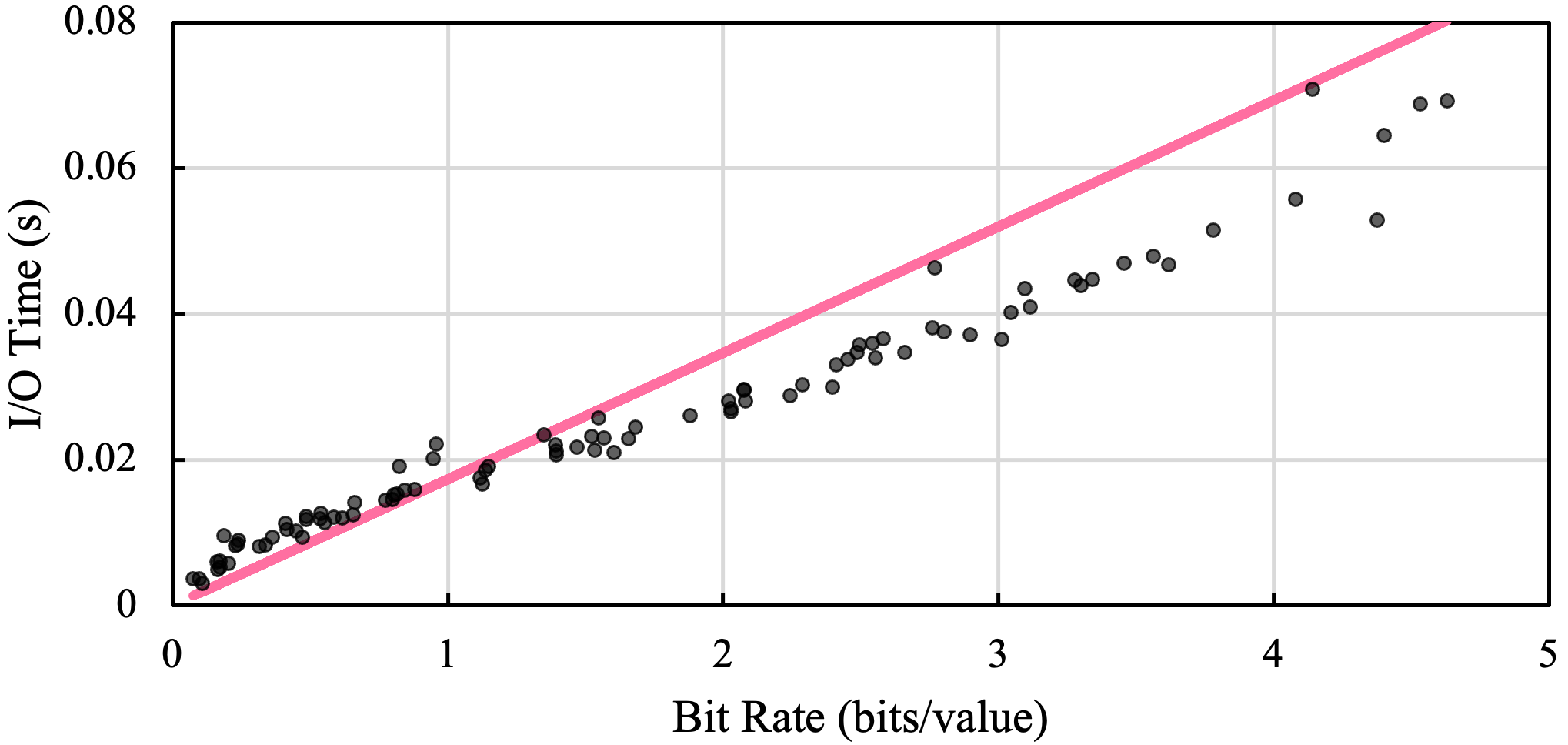}
    \caption{Accuracy of our write time estimation on $1024^3$ Nyx data samples. Red line is predicted time; black dots are actual time.
    }
    \label{fig:fig-thr-accuracy}
\end{figure}

\begin{figure*}[]\centering
\begin{subfigure}{0.49\linewidth}\centering
    \includegraphics[width=0.95\linewidth]{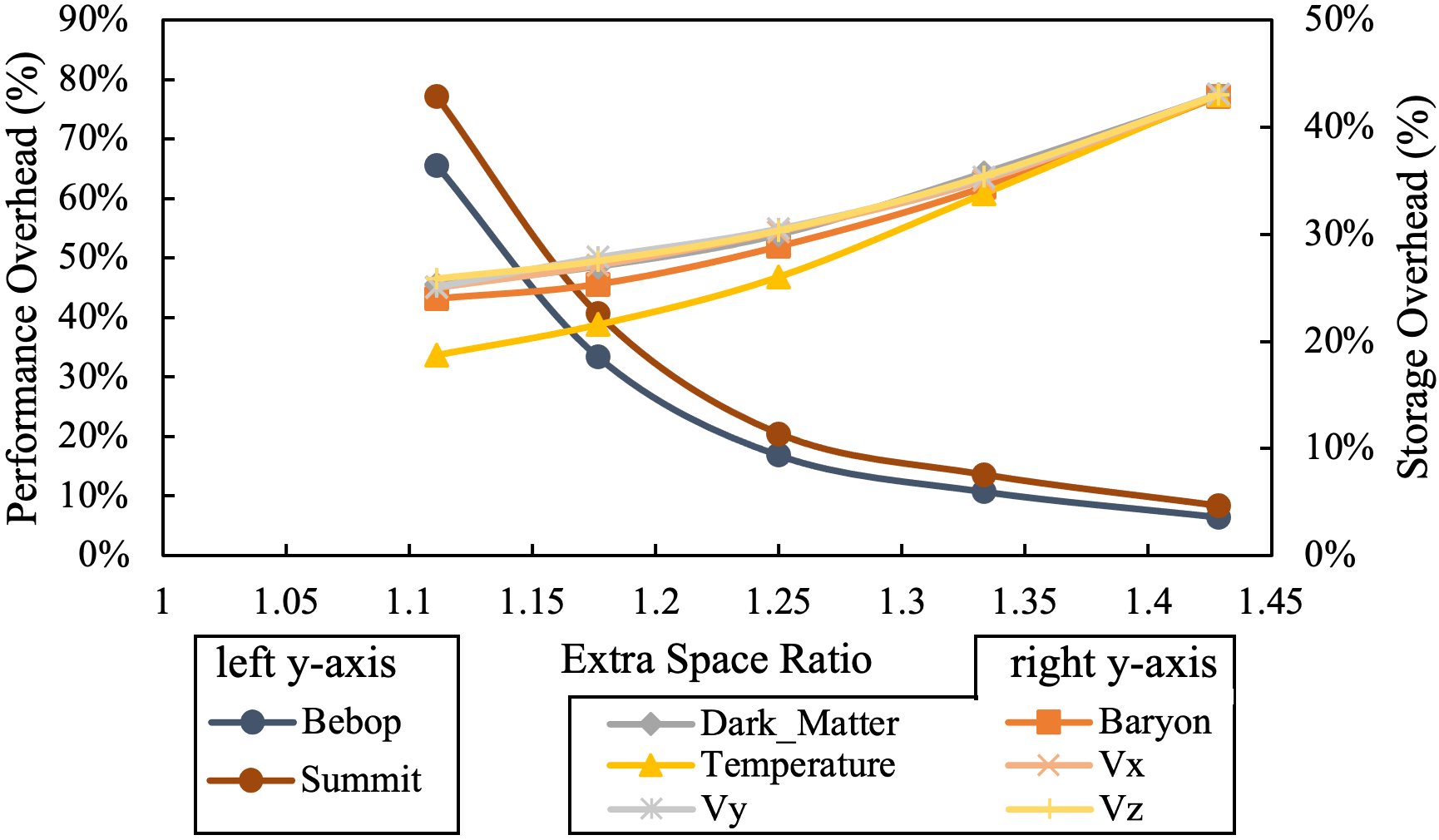}
    \vspace{-2mm}
	\caption{\footnotesize Nyx}\label{subfig:r1}
\end{subfigure}
\begin{subfigure}{0.49\linewidth}\centering
    \includegraphics[width=0.95\linewidth]{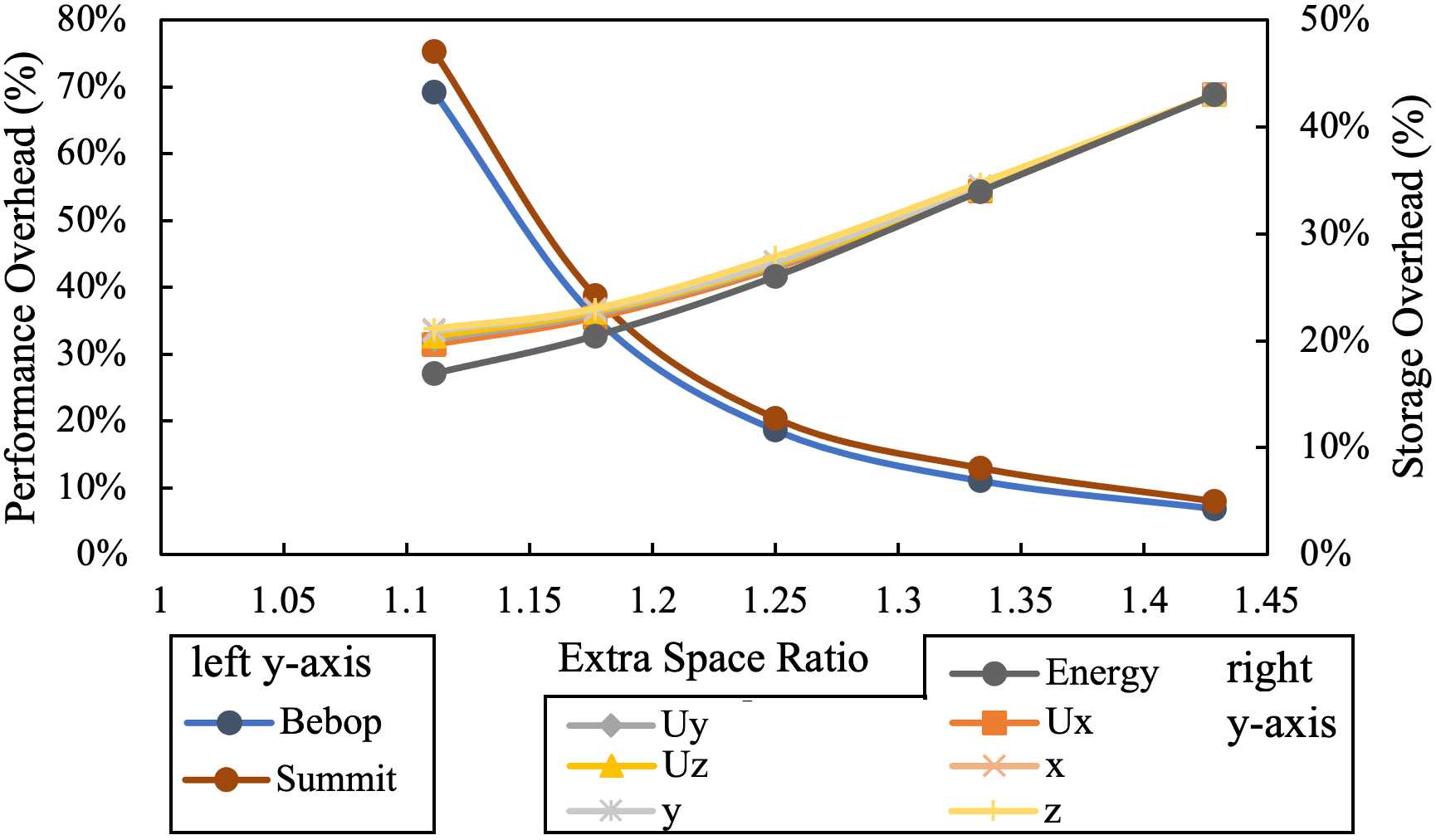}
    \vspace{-2mm}
	\caption{\footnotesize VPIC}\label{subfig:r2}
\end{subfigure}
\caption{Trade-off between performance overhead and storage overhead based on different extra space ratios on Nyx dataset (6 data fields) and VPIC dataset (7 data fields) on both Bebop and Summit with 512 processes. The target compressed bit-rate is 2 bits/value.}
\label{fig:fig-sec4-redundant}
\vspace{-6mm}
\end{figure*}

Similarly, we evaluate the accuracy of our proposed write-time estimation. 
We perform the offline evaluation by writing data from 128 processes into a shared file multiple times. The size of the data per process is set to 5 MBs, 10 MBs, 20 MBs, 50 MBs, or 100 MBs. We measure the average write throughput.
Based on our observation, further increasing or decreasing the scale would not significantly affect the average I/O throughput per process; therefore, we only perform the offline evaluation on one scale with different data sizes to reduce the offline evaluation overhead.
Next, we distribute the $1024^3$ Nyx dataset to 64 processes, perform the write operation, and measure the time of independent write for each process along with the bit-rate after compression.

Figure~\ref{fig:fig-thr-accuracy} shows the comparison between the estimated write time and the actual write time.
Note that the accuracy of low bit-rate is slightly lower than that of high bit-rate.
This is because the compressed data size is very small (i.e., 0.94 MB in this case) under low bit-rate, which result in a significantly low write throughput.
However, since we optimize the compression order based only on the absolute write time, such a small amount of write-time prediction error has hardly any effects on optimization decisions.

\subsection{Evaluation on Extra Space Ratio}
\label{sec:eval-redundant}

As mentioned in Section~\ref{sec:3-redundant}, the extra space ratio is one of the most important parameters that balance the overall write performance and storage overhead in our framework.
We then evaluate this performance-storage trade-off on both the Nyx and VPIC datasets.
For the Nyx dataset, we use the 
$1024^3$
dataset and distribute it to 512 processes for parallel write.
For the VPIC dataset, we downscale the original dataset by sampling 10\% data points and distribute it to 512 processes.

Figure~\ref{fig:fig-sec4-redundant} shows the correlation between the write-performance overhead and the storage overhead with the two datasets on both Bebop and Summit. Note that the write-performance overhead is compared with the write time without handling data overflow (excluding the compression time), and the storage overhead is compared with the ideal compressed data size (without the extra space).
We can observe that for each dataset on a given system, the trade-off curve is highly similar across different data fields, since the accuracy of the compression-ratio model is relatively stable on the same dataset.
We also note that even though the write-performance overhead and storage overhead with the same extra space ratio are different on the two datasets, the trade-off between the two overheads is very similar.
This is because the lower upper bound of the compression-ratio estimation results in larger number of processes to hold overflow data and hence higher performance overhead, which also enlarges the total amount of overflow data and increases the overall storage overhead.

\begin{figure}[]
    \centering
    \includegraphics[width=0.9\linewidth]{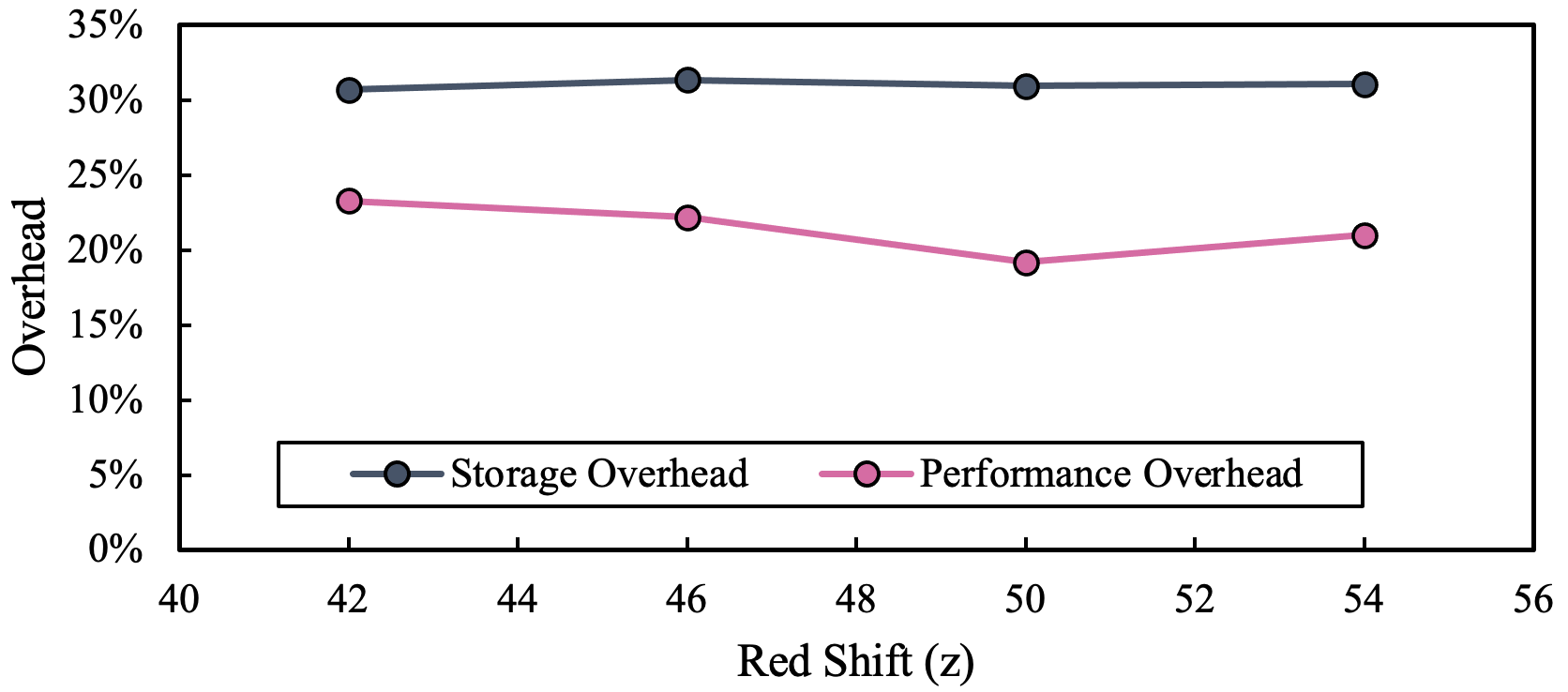}
    \caption{\revise{Evaluation on the consistency of the storage and performance overheads using the same extra space ratio of 1.25 with 512 processes on Summit. Red shift stands for different time-steps (higher values means earlier time in the simulation).}}
    \label{fig:fig-sec4-add-1}
\end{figure}

\begin{figure}[]
    \centering
    \includegraphics[width=\linewidth]{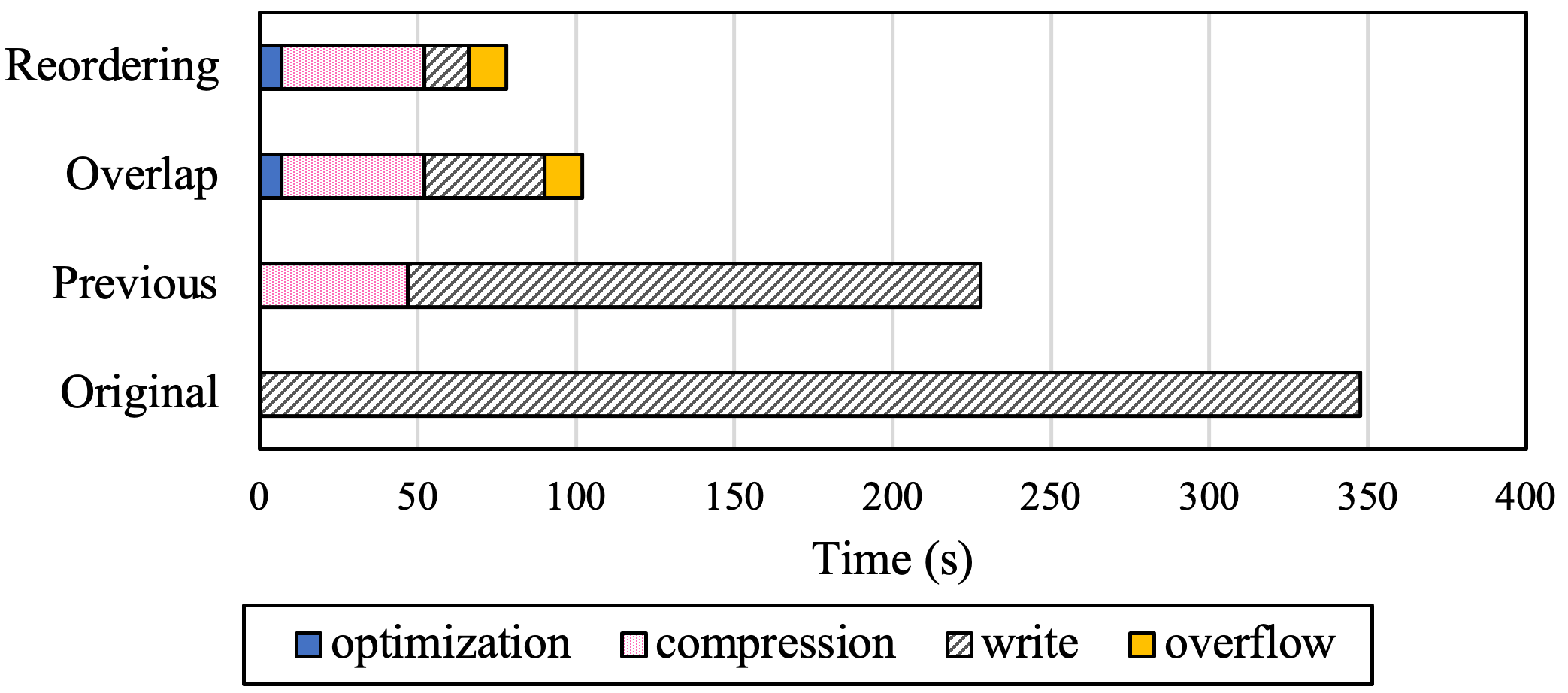}
    \caption{Performance comparison among our solution (overlapping and reordering), original non-compression solution, and previous compression-write solution on $4096^3$ Nyx dataset with 512 processes.}
    \label{fig:fig-sec4-overall-small}
\end{figure}

\begin{figure*}[]\centering
\begin{subfigure}{0.48\linewidth}\centering
    \includegraphics[width=0.99\linewidth]{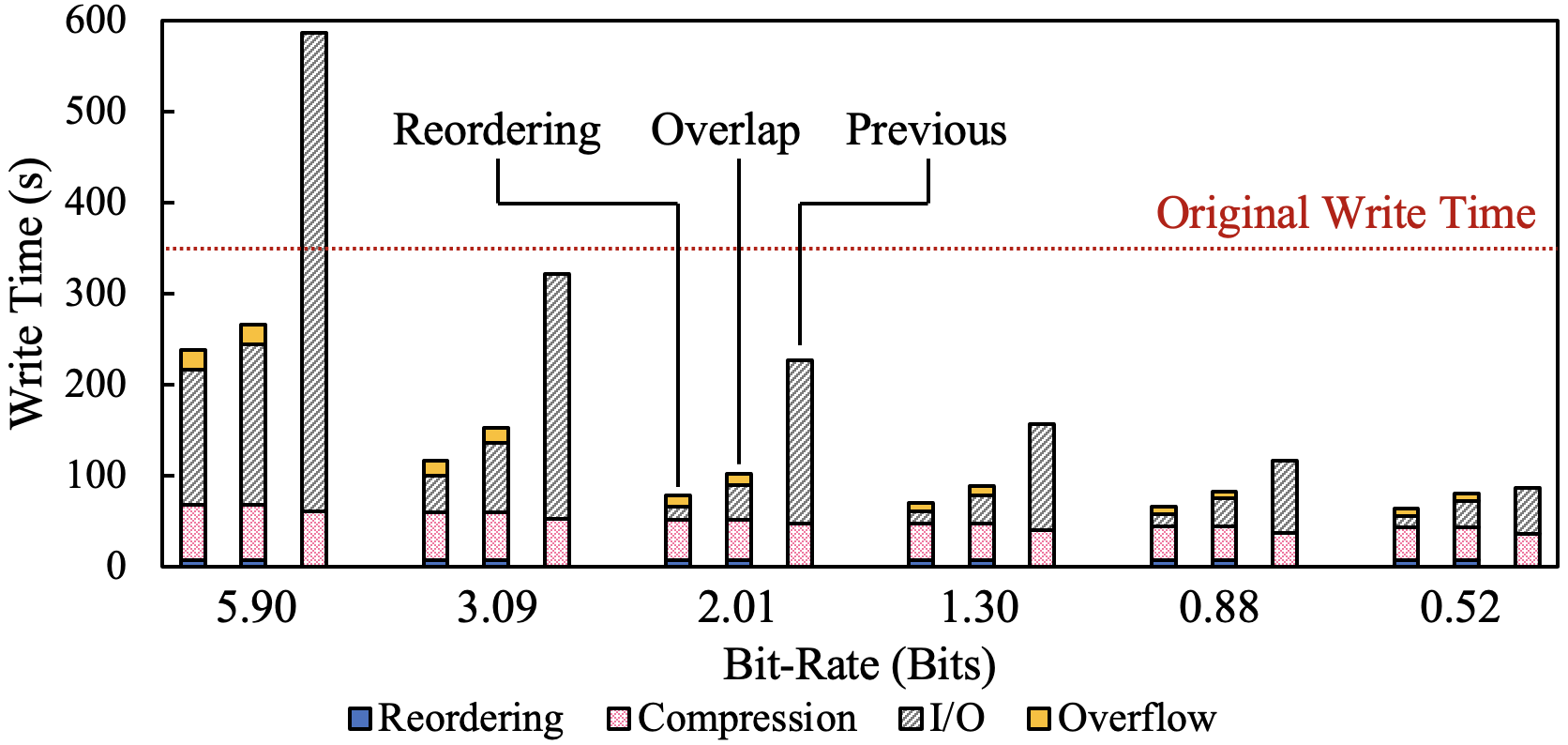}
	\caption{\footnotesize Nyx with different compression ratio}\label{subfig:o1}
\end{subfigure}
\begin{subfigure}{0.48\linewidth}\centering
    \includegraphics[width=0.99\linewidth]{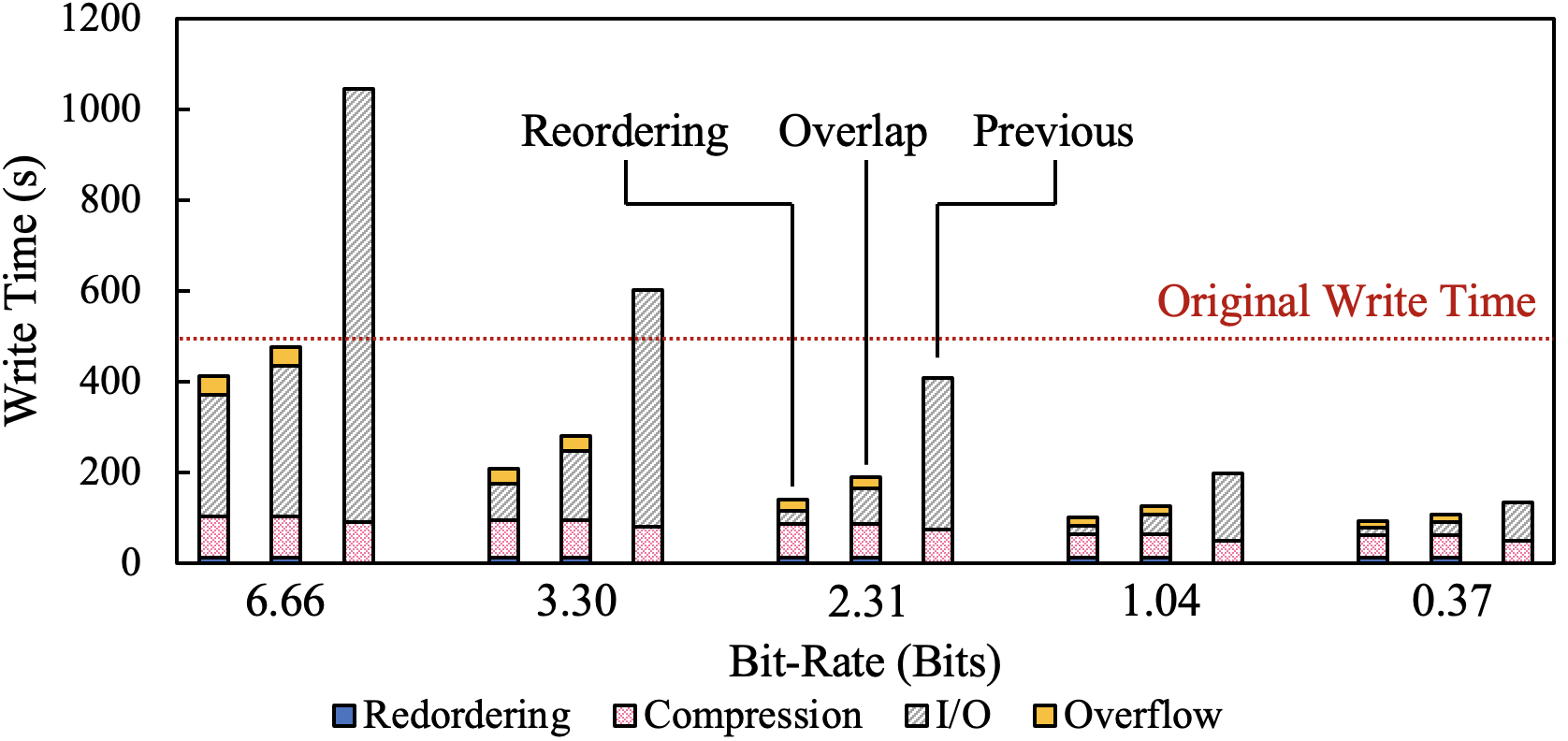}
	\caption{\footnotesize VPIC with different compression ratio}\label{subfig:o2}
\end{subfigure}
\begin{subfigure}{0.48\linewidth}\centering
    \includegraphics[width=0.99\linewidth]{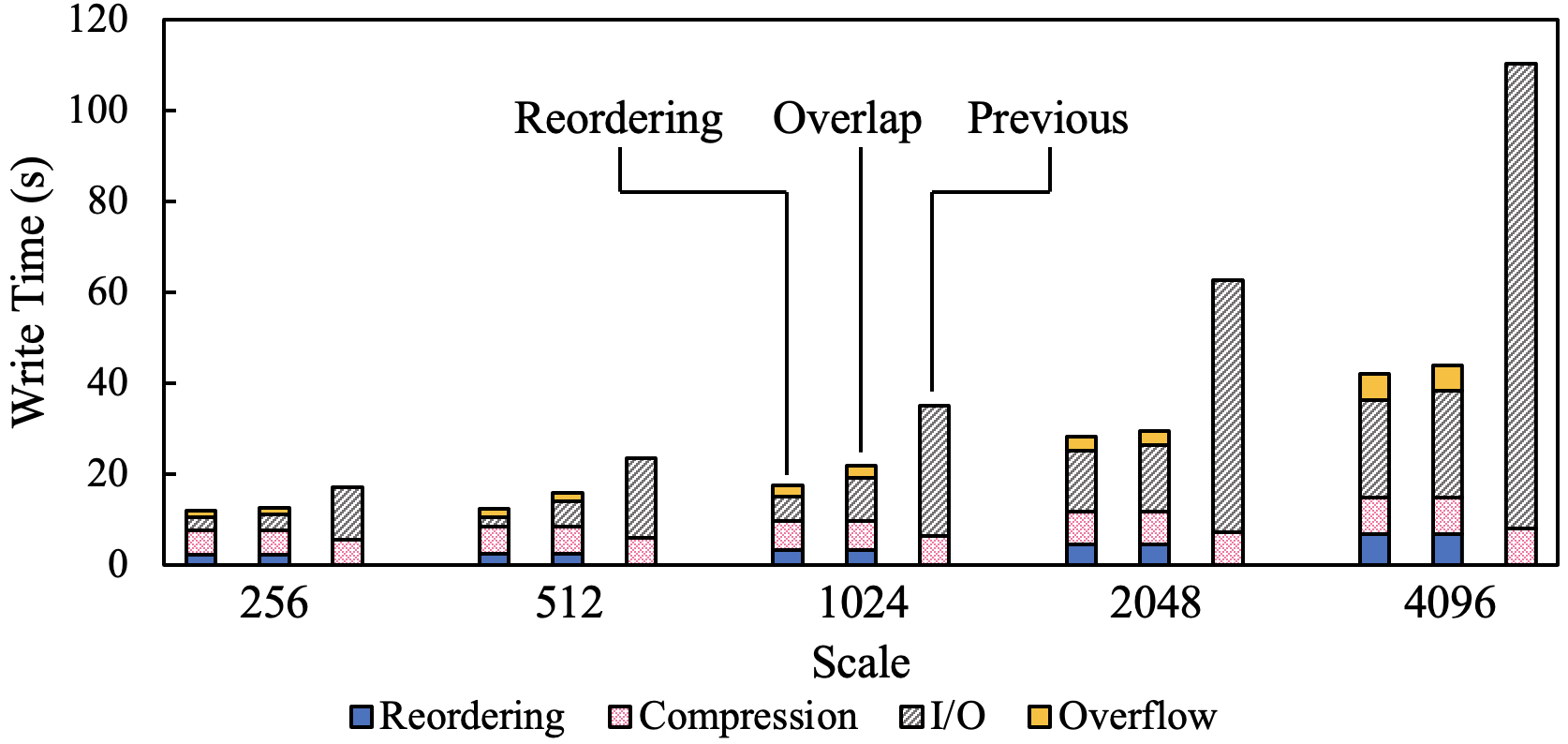}
	\caption{\footnotesize Nyx with different scale}\label{subfig:o3}
\end{subfigure}
\begin{subfigure}{0.48\linewidth}\centering
    \includegraphics[width=0.99\linewidth]{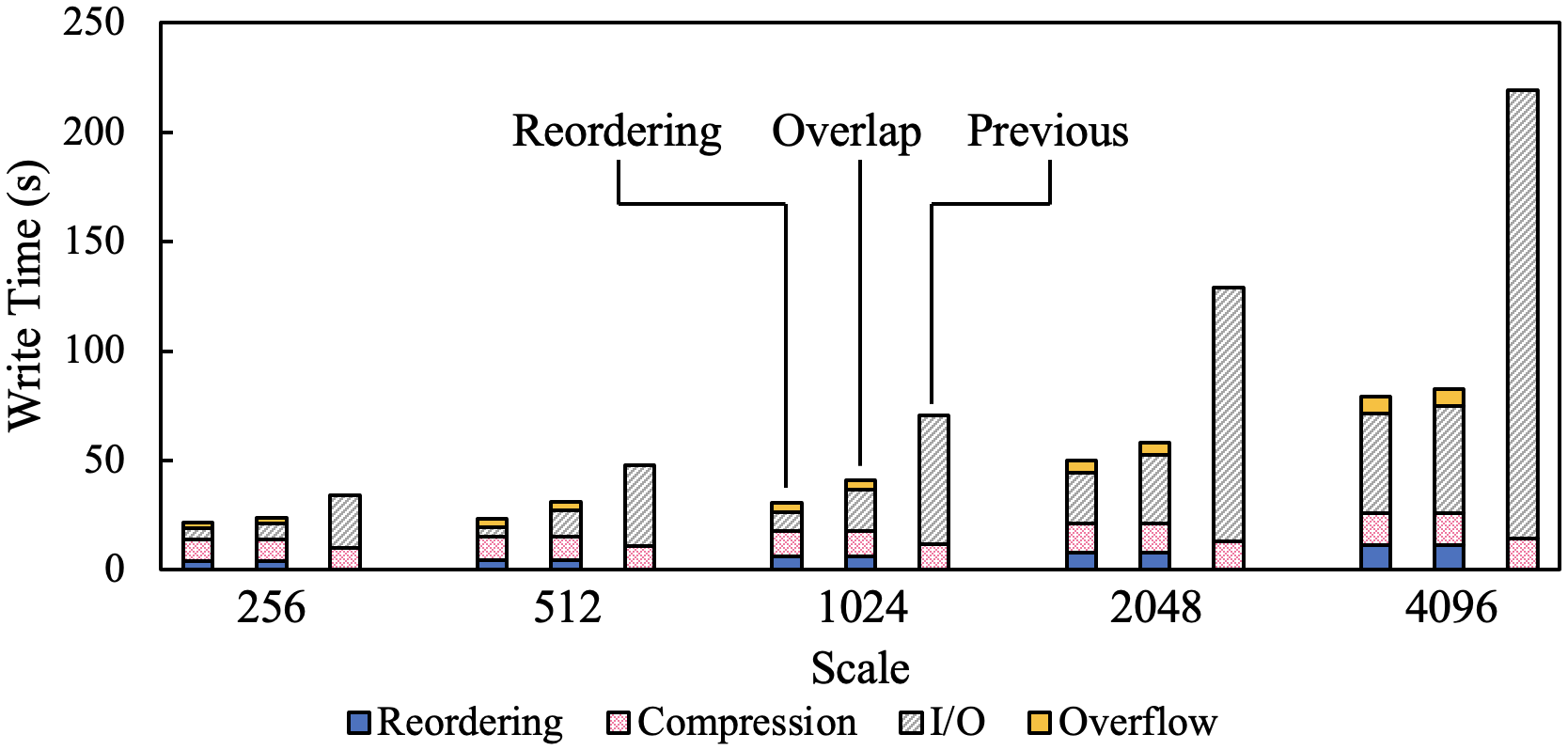}
	\caption{\footnotesize VPIC with different scale}\label{subfig:o4}
\end{subfigure}
\caption{\revise{Performance improvement of overall parallel-write with our proposed solution compared to the previous write solution with H5Z-SZ on both Nyx and VPIC datasets. Dashed red line is the baseline of HDF5 without compression. (a) and (b) are evaluated with 512 processes on Summit. (c) and (d) are evaluated with a target bit-rate of 2.}}
\label{fig:fig-sec4-overall}
\vspace{-6mm}
\end{figure*}

Regarding the results across the two systems, the difference of the performance overhead is mainly due to the higher I/O bandwidth of Summit over Bebop, which reduces the overall write time and enlarges the relative performance overhead.
However, the performance-storage tread-off is still similar across different systems, which means that we can use our offline study to guide the online decision making for the trade-off between performance and storage.
Figure~\ref{fig:fig-sec4-redundant} shows the averages of the performance and storage overhead in different extra space ratios across the four dataset-system setups (as mentioned in Section~\ref{sec:3-redundant}).
By default, we use the extra space ratio of 1.25 to minimize performance overhead while keeping low storage cost.
\revise{
In addition, we apply the extra space ratio of 1.25 with multiple time-steps in a series of Nyx datasets, as shown in Figure~\ref{fig:fig-sec4-add-1}.
It shows that our solution has consistent storage and performance overhead across time-steps.
Another alternative approach is to sample the dataset from the first time-step and provide users a more accurate estimation on the trade-off between performance and storage.
We will design a more user-friendly parameter-tuning mechanism in future.}

\subsection{Overall Performance Improvement}
\label{sec:eval-overall}

Next, we compare the performance of our proposed solution with the non-compression solution and the previous compression-filter solution.
For the non-compression solution, the distributed data is written to the shared file without any compression.
We implement it using HDF5 with independent write, which can significantly increase the performance over collective write~\cite{byna2020exahdf5}.
For the previous compression-filter solution, the compression process is considered as a filter, and the parallel write starts only when every process finishes its compression on all data partitions and the compressed data sizes are known to all processes to compute the write offsets.
We implement it using HDF5 with collective write, since compression filters do not support independent write.
Note that HDF5 only provides additional features such as asynchronous I/O and does not change the collective write.

\begin{figure*}[]\centering
\begin{subfigure}{0.48\linewidth}\centering
    \includegraphics[width=0.96\linewidth]{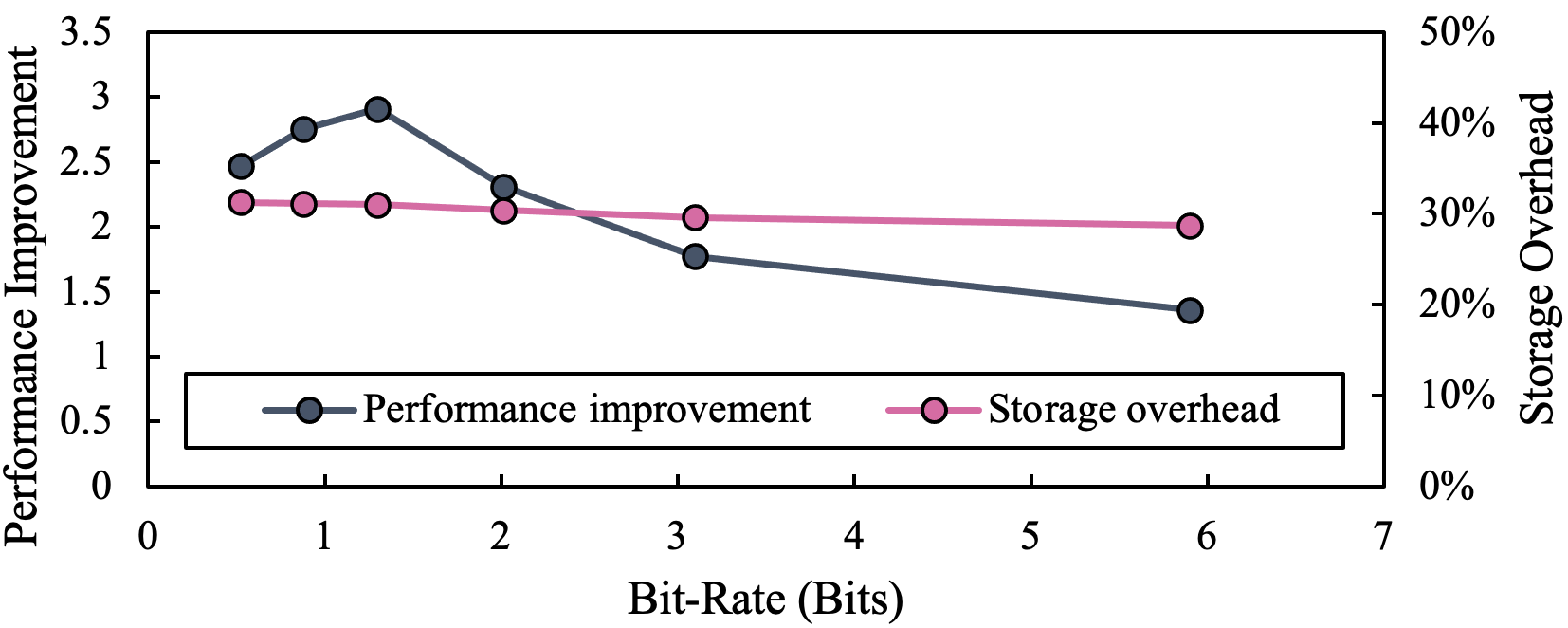}
	\caption{\footnotesize Nyx with different compression ratio}\label{subfig:o1-b}
\end{subfigure}
\begin{subfigure}{0.48\linewidth}\centering
    \includegraphics[width=0.96\linewidth]{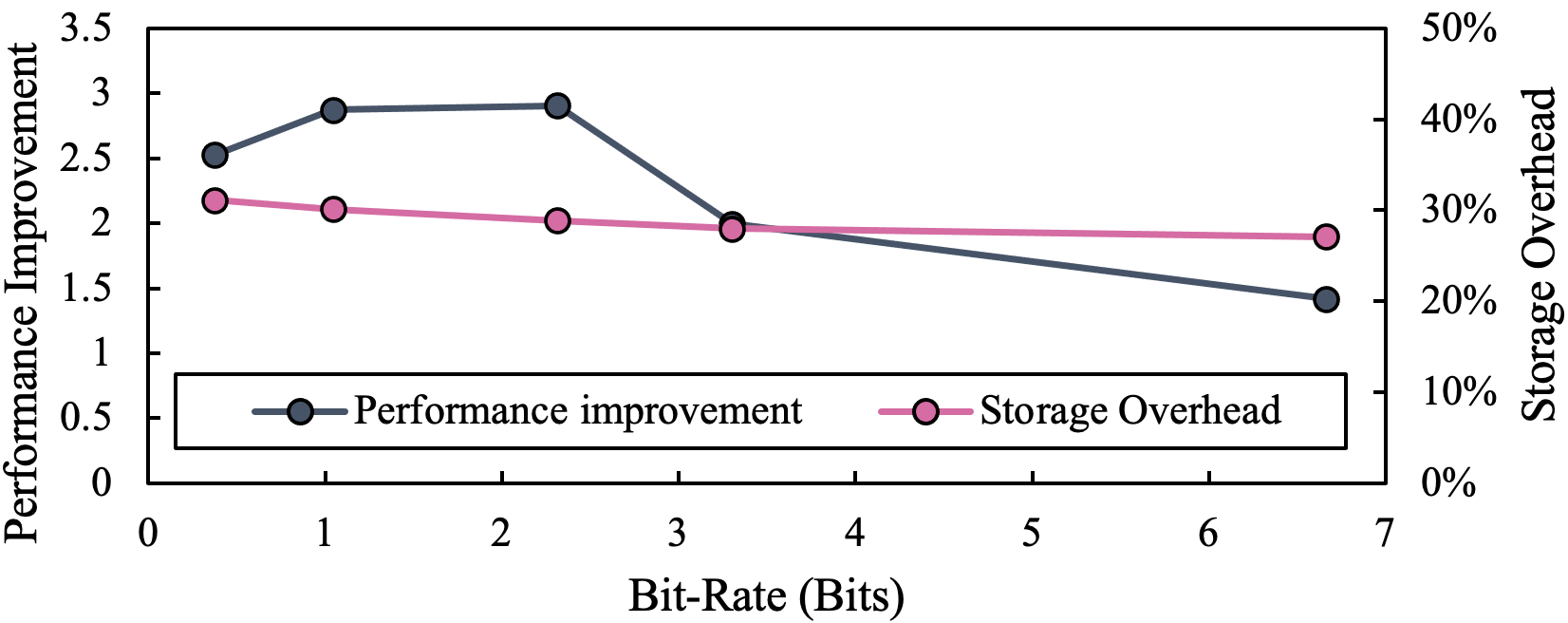}
	\caption{\footnotesize VPIC with different compression ratio}\label{subfig:o2-b}
\end{subfigure}
\begin{subfigure}{0.48\linewidth}\centering
    \includegraphics[width=0.96\linewidth]{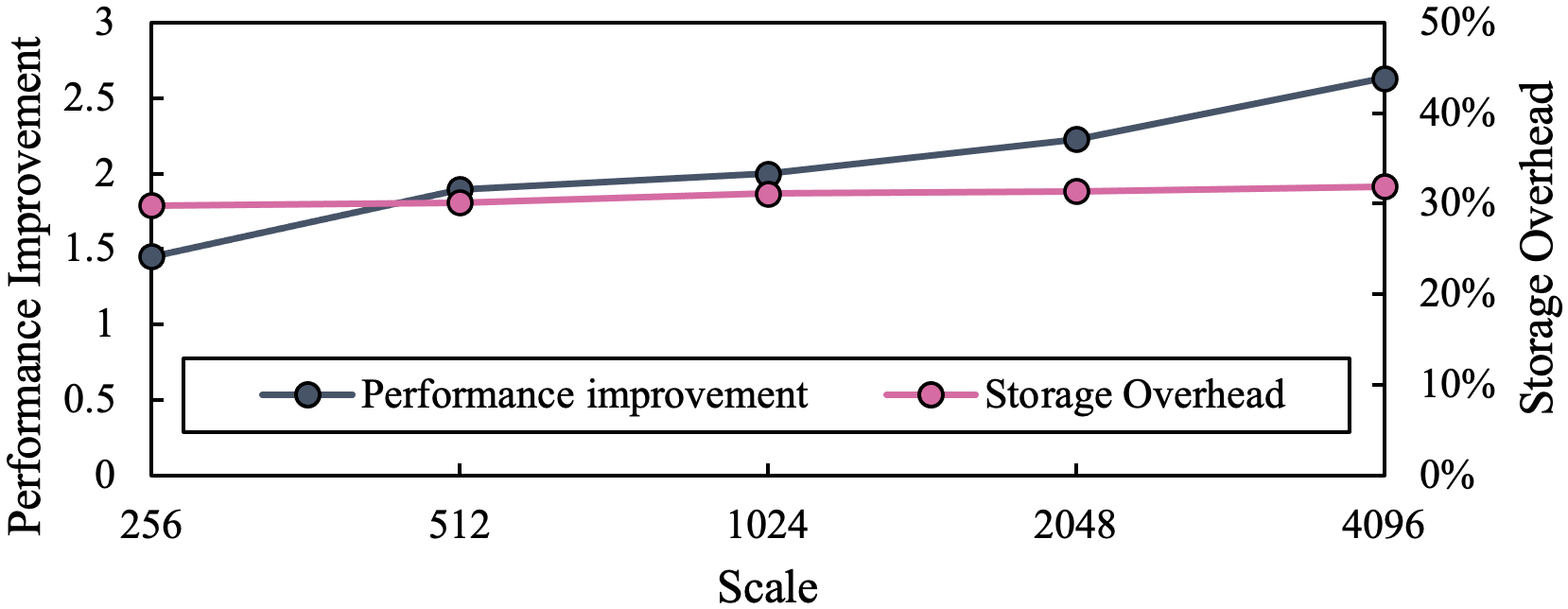}
	\caption{\footnotesize Nyx with different scale}\label{subfig:o3-b}
\end{subfigure}
\begin{subfigure}{0.48\linewidth}\centering
    \includegraphics[width=0.96\linewidth]{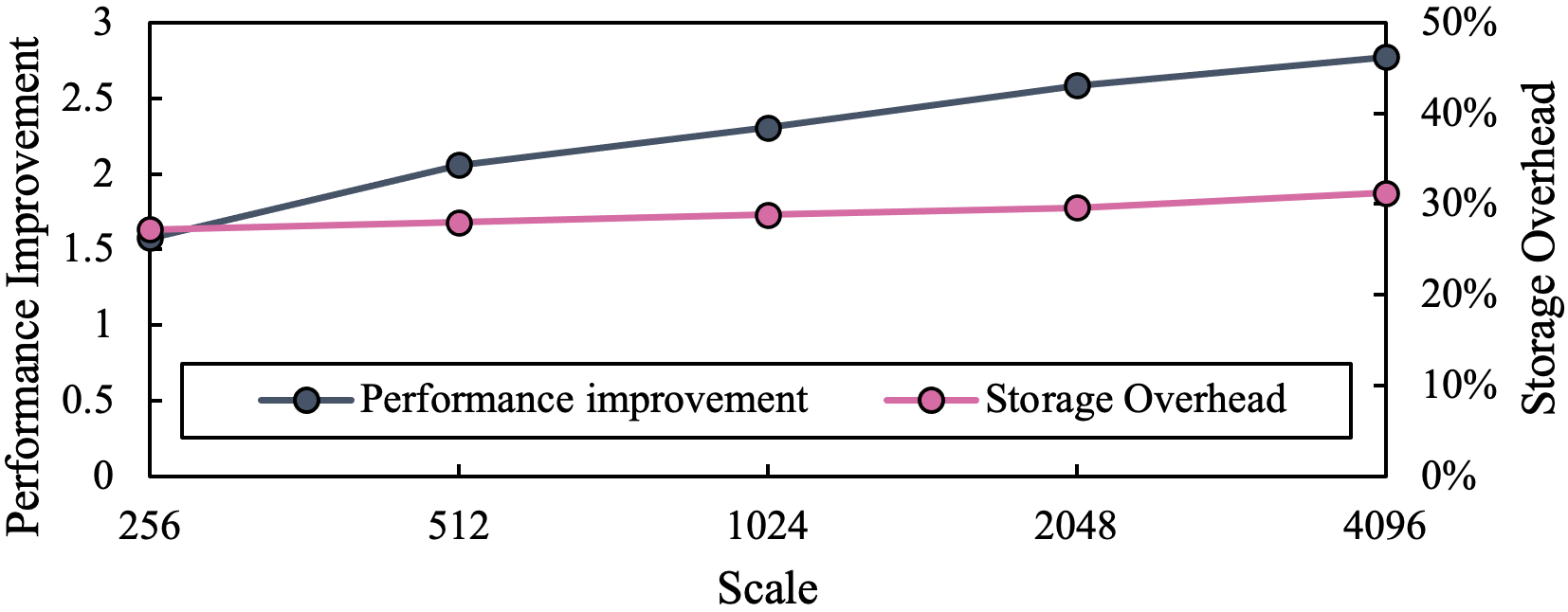}
	\caption{\footnotesize VPIC with different scale}\label{subfig:o4-b}
\end{subfigure}
\caption{\revise{Performance improvement (overall) and storage overhead of our solution compared to the previous solution on both Nyx and VPIC datasets. 
(a) and (b) are evaluated with 512 processes om Summit. (c) and (d) are evaluated with a target bit-rate of 2.}}
\label{fig:fig-sec4-overall}
\vspace{-6mm}
\end{figure*}

Figure~\ref{fig:fig-sec4-overall-small} shows the comparison of the performance breakdown between our proposed solution and the two existing solutions.
We implement our solution both with and without the compression reordering, referred to as ``overlapping'' and ``reordering'' in the figure, respectively.
Note that the write time bar shown in the overlapping solution is measured by the time between the end of the slowest compression and the end of the writing process rather than the entire write time (due to overlapping). 
The original/ideal compression ratio (without the extra space) in this experiment is $17.94\times$, while the actual compression ratio considering the extra space (i.e., the extra space ratio of  $1.25$) is $14.13\times$.

We can observe that the collective write solution with lossy compression still outperforms the independent write without compression by $1.87\times$ due to the high compression ratio and reduced data size.
Compared to the previous compression-write solution, our overlapping solution further improves the overall write performance by $1.79\times$ due to the asynchronous independent write.
Note that the compression times spent in the two solutions are similar, indicating that our framework improves the writing efficiency rather than the compression throughput.
Furthermore, with compression order optimization, we further improve the performance by $1.30\times$ in addition to our overlapping optimization.
The extra write time (gray bar) is significantly reduced due to the high overlapping efficiency.
Overall, comparing our optimized solution to the non-compression write, we improve the write performance by $4.46\times$ with the compression ratio of $14.1\times$; compared to the previous write with H5Z-SZ with the compression ratio of $17.9\times$, our optimized solution improves the write performance by $2.91\times$ with a $26\%$ storage overhead.
Note that this overhead is compared with the compressed data size, if compared with the original data size, the storage overhead is equivalent to $1.5\%$.

Furthermore, we evaluate our framework with different overall compression ratios and different scales to illustrate the efficiency of compression reordering on both the Nyx and VPIC datasets.
Figure~\ref{subfig:o1} and Figure~\ref{subfig:o2} show that the compression reordering optimization has a poor improvement over the overlapping-only solution under extremely high and low compression ratio,
\revise{the corresponding performance improvement and storage overhead are shown in Figure~\ref{subfig:o1-b} and Figure~\ref{subfig:o2-b}.}
For extremely high compression ratio, the compressed data size in each process is tiny, so the write time is significantly smaller than the compression time, as shown in Figure~\ref{subfig:o1} and Figure~\ref{subfig:o2} (the short gray bar versus the orange bar).
In this case, reordering the compression tasks provides little performance benefit, since the overall compression time cannot be further improved and the potential of reducing the extra write time is negligible.
For extremely low compression ratio, the compressed data size in each process is large, so the overall compression time is significantly shorter than the write time, where the overlapping efficiency is likely to be sufficiently high even without optimization (low potential to optimize).
Both datasets have a similar compression ratio ($14.1\times$ and $10.9\times$ for Nyx and VPIC, respectively) that benefits the compression reordering optimization.
Note that although the higher compression ratio almost always indicates the better write performance, it also means lower data quality with information loss that may potentially hurt post-hoc analysis.
Under the extreme cases when the compression ratio is very small, the previous write solution with H5Z-SZ shows even worse performance than the non-compression write.

Finally, we perform the scaling study of our proposed framework. Specifically, we conduct a weak scaling study that keeps the same data partition size in each process, i.e., 
$256^3$
and $39,379,260$ for the Nyx and the VPIC datasets, respectively. 
Figure~\ref{subfig:o3} and Figure~\ref{subfig:o4} illustrate that the compression order optimization over the overlapping solution is relatively stable on the evaluated scales at $(256, 512, 1024, 2048, 4096)$,
\revise{the corresponding performance improvement and storage overhead are shown in Figure~\ref{subfig:o3-b} and Figure~\ref{subfig:o4-b}.}
This is because the overall compression time and write time in each process are mostly stable through different scales.
However, a larger scale slightly benefits our solution because the independent, asynchronous write typically provides better scalability compared to the collective write used by the previous compression-write solution~\cite{byna2020exahdf5}.
Note that the optimization time and the overflow time increase with the scale, because even though the prediction time is stable, larger scale introduces longer communication time for the all-gather operation.

In conclusion, under the circumstances of satisfying the user-set compression configuration, our parallel write with lossy compression is scalable and high-performance, especially when (1) the number of data fields is relatively large; (2) the overall compression ratio is preferably between $10\times$ and $20\times$ (with balanced compression time and write time); and (3) the data amount in each process is not too small (deserving compression). 
In addition, users can fine-tune the extra space ratio to make a good performance-storage trade-off.

\section{Conclusion and Future Work}
\label{sec:conclusion}

In this paper, we propose to integrate predictive lossy compression deeply with HDF5 to significantly improve the parallel-write performance for large-scale scientific simulations. 
We introduce a newly designed prediction method to estimate the compression and write time for each process and use this information to pre-compute the write offset before the actual compression.
Furthermore, we introduce an extra space design to handle the uncertainty of the prediction and an optimization algorithm to reorder the compression tasks in each process.
We evaluate our proposed solution on both Bebop cluster and Summit supercomputer with up to 4,096 cores.
The evaluation shows that our solution can provide a $4.46\times$ performance improvement with a $14.1\times$ compression ratio compared to the original parallel write without compression, and provide a $2.91\times$ performance improvement compared to the previous parallel write with the SZ compression filter with only $20\%$ compression ratio degradation.

In the future, we will extend our solution to other parallel I/O libraries such as ADIOS~\cite{godoy2020adios} and support more lossy compressors such as ZFP~\cite{zfp}.
\revise{
Moreover, we plan to evaluate our solution on more 
real-world HPC datasets. 
}
\section*{Acknowledgments}

\small This research was supported by the Exascale Computing Project (ECP), Project Number: 17-SC-20-SC, a collaborative effort of two DOE organizations---the Office of Science and the National Nuclear Security Administration, responsible for the planning and preparation of a capable exascale ecosystem, including software, applications, hardware, advanced system engineering and early testbed platforms, to support the nation's exascale computing imperative. The material was supported by the U.S. Department of Energy, Office of Science, Advanced Scientific Computing Research (ASCR), under contracts DE-AC02-06CH11357 and DE-AC02-05CH11231. This work was also supported by the National Science Foundation under Grants OAC-2003709, OAC-2042084, OAC-2104023, and OAC-2104024. We gratefully acknowledge the computing resources provided on Bebop cluster at Argonne
and Summit supercomputer at Oak Ridge.

\bibliographystyle{IEEEtran}
\bibliography{refs}

\end{document}